\documentclass{emulateapj}
\usepackage{psfig,enumerate}
\usepackage{color}
\usepackage{atbegshi}
\def\hackaltaffiltext#1#2{\AtBeginShipoutNext{\footnotetext[#1]{#2}\stepcounter{footnote}}}

\submitted{}

\newcommand{\kms}{\mbox{km s$^{-1}~$}}

\newcommand{\msune}{M$_{\odot}$}

\newcommand{\dgr}{$^{\circ}~$}

\newcommand{\teff}{$T_{\rm eff}$~}
\newcommand{\teffe}{$T_{\rm eff}$}
\newcommand{\logg}{$\log{g}$~}

\newcommand{\hie}{\ion{H}{1}}

\begin{document}

\title{Tracing chemical evolution over the extent of \\ the Milky Way's Disk with APOGEE Red Clump Stars}

\shorttitle{APOGEE RC $\alpha$-abundances}
\shortauthors{NIDEVER ET AL.}

\author{David L. Nidever\altaffilmark{1},
Jo Bovy\altaffilmark{2,3},
Jonathan C. Bird\altaffilmark{4},
Brett H. Andrews\altaffilmark{5},
Michael Hayden\altaffilmark{6},
Jon Holtzman\altaffilmark{6},
Steven R. Majewski\altaffilmark{7},
Verne Smith\altaffilmark{8},
Annie C. Robin\altaffilmark{9},
Ana E. Garc{\'{\i}}a P{\'e}rez\altaffilmark{7},
Katia Cunha\altaffilmark{10,11},
Carlos Allende Prieto\altaffilmark{12,13},
Gail Zasowski\altaffilmark{14},
Ricardo P. Schiavon\altaffilmark{15},
Jennifer A. Johnson\altaffilmark{5},
David H. Weinberg\altaffilmark{5},
Diane Feuillet\altaffilmark{6},
Donald P. Schneider\altaffilmark{16,17},
Matthew Shetrone\altaffilmark{18},
Jennifer Sobeck\altaffilmark{9},
D. A. Garc\'ia-Hern\'andez\altaffilmark{12,13},
O. Zamora\altaffilmark{12,13},
Hans-Walter Rix\altaffilmark{19},
Timothy C. Beers\altaffilmark{20,21},
John C. Wilson\altaffilmark{7},
Robert W. O'Connell\altaffilmark{7},
Ivan Minchev\altaffilmark{22},
Cristina Chiappini\altaffilmark{22,23},
Friedrich Anders\altaffilmark{22},
Dmitry Bizyaev\altaffilmark{24},
Howard Brewington\altaffilmark{24},
Garrett Ebelke\altaffilmark{24},
Peter M. Frinchaboy\altaffilmark{25},
Jian Ge\altaffilmark{26},
Karen Kinemuchi\altaffilmark{24},
Elena Malanushenko\altaffilmark{24},
Viktor Malanushenko\altaffilmark{24},
Moses Marchante\altaffilmark{24},
Szabolcs~M{\'e}sz{\'a}ros\altaffilmark{27,28},
Daniel Oravetz\altaffilmark{24},
Kaike Pan\altaffilmark{24},
Audrey Simmons\altaffilmark{24},
Michael F. Skrutskie\altaffilmark{7}
}


\altaffiltext{1}{Department of Astronomy, University of Michigan,
Ann Arbor, MI, 48104, USA (dnidever@umich.edu)}

\altaffiltext{2}{Institute for Advanced Study, Einstein Drive, Princeton, NJ 08540, USA}

\altaffiltext{3}{Hubble Fellow}

\altaffiltext{4}{Physics and Astronomy Department, Vanderbilt University, 1807 Station B, Nashville, TN 37235, USA}

\altaffiltext{5}{Department of Astronomy and the Center for Cosmology and
Astro-Particle Physics, The Ohio State University, Columbus, OH 43210, USA}

\altaffiltext{6}{New Mexico State University, Las Cruces, NM 88003, USA}

\altaffiltext{7}{Department of Astronomy, University of Virginia, Charlottesville, VA, 22904, USA}

\altaffiltext{8}{National Optical Astronomy Observatory, Tucson, AZ 85719, USA}

\altaffiltext{9}{Institut Utinam, CNRS UMR 6213, OSU THETA, Universit{\'e} de Franche-Comt{\'e},
41bis avenue de l'Observatoire, 25000 Besan\c{c}on, France}

\altaffiltext{10}{Observatorio Nacional, Rio de Janeiro, Brazil}

\altaffiltext{11}{Steward Observatory, 933 N. Cherry St., University of Arizona, Tucson, AZ 85721, USA}

\altaffiltext{12}{Instituto de Astrofísica de Canarias, E38205 La Laguna, Tenerife, Spain}

\hackaltaffiltext{13}{Departamento de Astrofísica, Universidad de La Laguna (ULL), E-38206 La Laguna, Tenerife, Spain}

\hackaltaffiltext{14}{Department of Physics and Astronomy, Johns Hopkins University, Baltimore, MD 21218, USA}

\hackaltaffiltext{15}{Astrophysics Research Institute, IC2, Liverpool Science Park, Liverpool John Moores
University, 146 Brownlow Hill, Liverpool, L3 5RF, UK}

\hackaltaffiltext{16}{Department of Astronomy and Astrophysics, The Pennsylvania State University,
   University Park, PA 16802}

\hackaltaffiltext{17}{Institute for Gravitation and the Cosmos, The Pennsylvania State University,
   University Park, PA 16802}

\hackaltaffiltext{18}{University of Texas at Austin, McDonald Observatory, 32 Fowlkes Rd.,
McDonald Observatory, Tx 79734-3005}

\hackaltaffiltext{19}{Max-Planck-Institut f\"ur Astronomie, K\"onigstuhl 17, D-69117 Heidelberg, Germany}

\hackaltaffiltext{20}{Department of Physics, University of Notre Dame, 225 Nieuwland Science Hall, Notre Dame, IN 46656, USA}

\hackaltaffiltext{21}{JINA: Joint Institute for Nuclear Astrophysics, University of Notre Dame, Notre Dame, IN 46556, USA}

\hackaltaffiltext{22}{Leibniz-Institut f\"ür Astrophysik Potsdam (AIP), An der Sternwarte 16, 14482 Potsdam, Germany}

\hackaltaffiltext{23}{Laborat\'orio Interinstitucional de e-Astronomia, - LIneA, Rua Gal. Jos\'e Cristino 77,
Rio de Janeiro, RJ - 20921-400, Brazil}

\hackaltaffiltext{24}{Apache Point Observatory and New Mexico State
University, P.O. Box 59, Sunspot, NM, 88349-0059, USA}

\hackaltaffiltext{25}{Department of Physics \& Astronomy, Texas Christian University, Fort Worth, TX, 76129}

\hackaltaffiltext{26}{Astronomy Department, University of Florida, Gainesville, FL 32611-2055}

\hackaltaffiltext{27}{Department of Astronomy, Indiana University, Bloomington, IN 47405, USA}

\hackaltaffiltext{28}{ELTE Gothard Astrophysical Observatory, H-9704 Szombathely, Szent Imre herceg st. 112, Hungary}

\begin{abstract}
We employ the first two years of data from the near-infrared, high-resolution SDSS-III/APOGEE
spectroscopic survey to investigate the distribution of metallicity
and $\alpha$-element abundances of stars over a large part of the Milky Way disk.
Using a sample of $\approx10,000$ kinematically-unbiased red-clump stars with $\sim$5\% distance accuracy
as tracers, the [$\alpha$/Fe]~vs.~[Fe/H] distribution of this sample exhibits a bimodality in [$\alpha$/Fe] at
intermediate metallicities, $-0.9<$[Fe/H]$<-0.2$, but at higher
metallicities ([Fe/H]$\sim$+0.2) the two sequences smoothly merge.
We investigate the effects of the APOGEE 
selection function and volume filling fraction and find that these have
little qualitative impact on the $\alpha$-element abundance patterns. The
described abundance pattern is found throughout the range
5$<$$R$$<$11 kpc and 0$<$$|Z|$$<$2 kpc across the Galaxy.
The [$\alpha$/Fe] trend of the high-$\alpha$ sequence is surprisingly
constant throughout the Galaxy, with little variation from region to region ($\sim$10\%).
Using simple galactic chemical evolution models we derive an average star formation efficiency
(SFE) in the high-$\alpha$ sequence of $\sim$4.5$\times$10$^{-10}$~yr$^{-1}$,
which is quite close to the nearly-constant value found in molecular-gas-dominated regions of nearby spirals.
This result suggests that the early evolution of the Milky Way disk was characterized by stars
that shared a similar star formation history and were formed in a well-mixed, turbulent, and
molecular-dominated ISM with a gas consumption timescale (SFE$^{-1}$) of $\sim 2$ Gyr.
Finally, while the two $\alpha$-element sequences in the inner Galaxy can be explained by a single chemical
evolutionary track this cannot hold in the outer Galaxy,
requiring instead a mix of two or more populations with distinct enrichment histories.
\end{abstract}

\keywords{
  Galaxy: abundances --- 
  Galaxy: disk ---
  Galaxy: evolution ---
  Galaxy: stellar content ---
  Galaxy: structure --- 
  surveys}

\section{Introduction}
\label{sec:intro}

The Milky Way galaxy (MW) is a cornerstone in the study of the
internal structure and evolution of large disk galaxies, because stellar
populations in the MW can be studied using detailed observations of
large samples of individual stars \citep[e.g.,][]{Rix13a}. Over the past few
decades these surveys have led to the discovery of disk stars at large distances
above the plane through star counts (the thick disk;
\citealt{Yoshii82,Gilmore83}), observations of abundance gradients over the
extent of the disk
\citep[e.g.,][]{Audouze76a,Chen03a,Allende06,Cheng12a,Boeche14}, and a
detailed mapping of the local distribution of elemental abundances
\citep[e.g.,][]{vandenbergh62a,Fuhrmann98,Adibekyan12}.
In addition, the first year of high-resolution spectroscopy from the Sloan
Digital Sky Survey III's Apache Point Observatory Galactic Evolution Experiment
(APOGEE) was used by \citet{Hayden14a} and \citet{Anders14} to study the
Milky Way's disk over a large area.  These
measurements provide crucial constraints on models for the formation
and evolution of the MW disk \citep[e.g.,][]{Larson76a,Chiappini97}.
In this paper we extend these observations by tracing the detailed
distribution of elemental abundances of red clump stars over a large part
of the MW's disk using the first two years of APOGEE data.

In the solar neighborhood, the distribution of stars in the
([$\alpha$/Fe],[Fe/H]) plane displays a sequence extending from
metal-poor, high-$\alpha$ stars to about solar abundances that is believed to be 
associated with the thick disk, because of the large velocity dispersion of its stars 
\citep{Fuhrmann98,Prochaska00,Bensby05,Reddy06,Ramirez07,Lee11}. This sequence was
originally established by observing stars kinematically selected to be likely
members of a thick-disk population; this kinematical bias has made
the interpretation of this sequence, its precise relation to the thick disk, and its extension
to solar abundances difficult \citep[e.g.,][]{Bensby07}. Recently, the
kinematically-unbiased HARPS sample \citep{Adibekyan12} has removed
this obstacle, and demonstrated that the high-$\alpha$ stars extend to solar and
super-solar metallicities (\citealt{Adibekyan11,Adibekyan13}; see
also \citealt{Bensby14}). The stars in this high-$\alpha$ sequence
have ages of $\sim$7 Gyr and larger \citep{Haywood13}, with more $\alpha$-enhanced objects being older,
with ages up to 12 Gyr \citep{Bensby05}. The spread in age and
[$\alpha$/Fe] for stars in the thick disk indicates an extended
star-formation history with time for enrichment by Type Ia supernovae
(a few Gyr; \citealt{Maoz11a}).

Recently, progress has been made in mapping the spatial distribution
of chemically differentiated stellar populations in the disk. In particular, it has become
clear that high-$\alpha$ stars have a shorter radial scale length than stars with solar [$\alpha$/Fe]
\citep{Bensby11a,Bovy12b,Cheng12b,Anders14}. \citet{Bovy12b} mapped the spatial
distribution (scale height and length) of different mono-abundance
populations in detail, finding a complex dependence of the radial
scale length on ([$\alpha$/Fe],[Fe/H]), a smooth distribution of scale-heights
ranging between 200 pc and 1 kpc, and a similarly smooth increase of
the velocity dispersion with [$\alpha$/Fe] (\citealt{Bovy12a,Bovy12c};
see also \citealt{Haywood13}). After accounting for the spatial selection function of SEGUE, 
Bovy et al.\ find a smooth distribution in the ([$\alpha$/Fe],[Fe/H]) plane at
the solar cylinder, finding no distinct gap along the
high-$\alpha$ sequence seen in other studies
\citep[e.g.,][]{Fuhrmann98,Reddy06,Adibekyan11}. This behavior can be interpreted as showing
that the scale-height of the MW disk continuously and gradually
decreased over time as the disk was enriched with metals, while
the [$\alpha$/Fe] abundances decreased (due to Type Ia supernovae
becoming the dominant form of enrichment). The complicated
radial behavior of different mono-abundance populations---with the short
scale lengths for the high-$\alpha$ stars, longer scale lengths for stars
with solar abundances, and almost constant radial densities for low-$\alpha$,
low-[Fe/H] stars---means, however, that it is difficult to extrapolate the
local abundance distribution in order to further constrain the radial
profiles. It is therefore essential to trace the ([$\alpha$/Fe],[Fe/H])
distribution over a much wider range of Galactocentric radii to obtain a full
picture of the large-scale chemical structure of the disk.

Various qualitatively different scenarios for the formation of the
thick, old, high-$\alpha$ component in the MW have been suggested since the result
was first identified by \citet{Yoshii82} and \citet{Gilmore83}.  Many of these
ideas were proposed in the first decade after the discovery and were reviewed by, for
example, \citet{Majewski93a}. Some of these mechanisms rely on external events
such as satellite heating, accretion, or merger-induced star formation
\citep[e.g.,][]{Abadi03a,Quinn93a,Brook04a}, while others rely on internal evolution of
the disk through slow or more rapid dissipational collapse during disk
formation \citep{Larson76a,Gilmore84a} or secular disk
heating. A qualitatively new thick-disk formation
scenario, suggested more recently, is stellar radial migration via spiral wave scattering at
corotational resonance \citep{Sellwood02a}. \citet{Schoenrich09b} suggested that this mechanism is
able to produce a thick-disk component in agreement with many local observations, although more
realistic simulations cast serious doubt on whether radial migration can produce enough
\citep{Roskar13a} or even any \citep{Minchev12b,Vera-Ciro14b} heating.
The results of \citet{Bovy12b} disfavor an external origin for the thick disk, and
find that the local observations
can be reproduced by models where radial migration plays a large
role.  However, the observations can also be
explained by models where the thick disk formed largely as it is seen
today, from a hot interstellar medium at the onset of disk formation,
as favored by recent cosmological simulations
\citep[e.g.,][]{Bird13,Stinson13} and observations of high-mass disk
galaxies at $z \approx 2$ \citep[e.g.,][]{ForsterSchreiber11a}, or by
a combination of early merging with radial migration
\citep{Minchev13}. A detailed mapping of the chemical structure of the disk away
from the solar neighborhood will allow insights into the evolution of different
regions of the MW and provide qualitatively new constraints on the evolutionary
models described above.

\begin{figure}[ht!]
\begin{center}
\includegraphics[angle=0,scale=0.55]{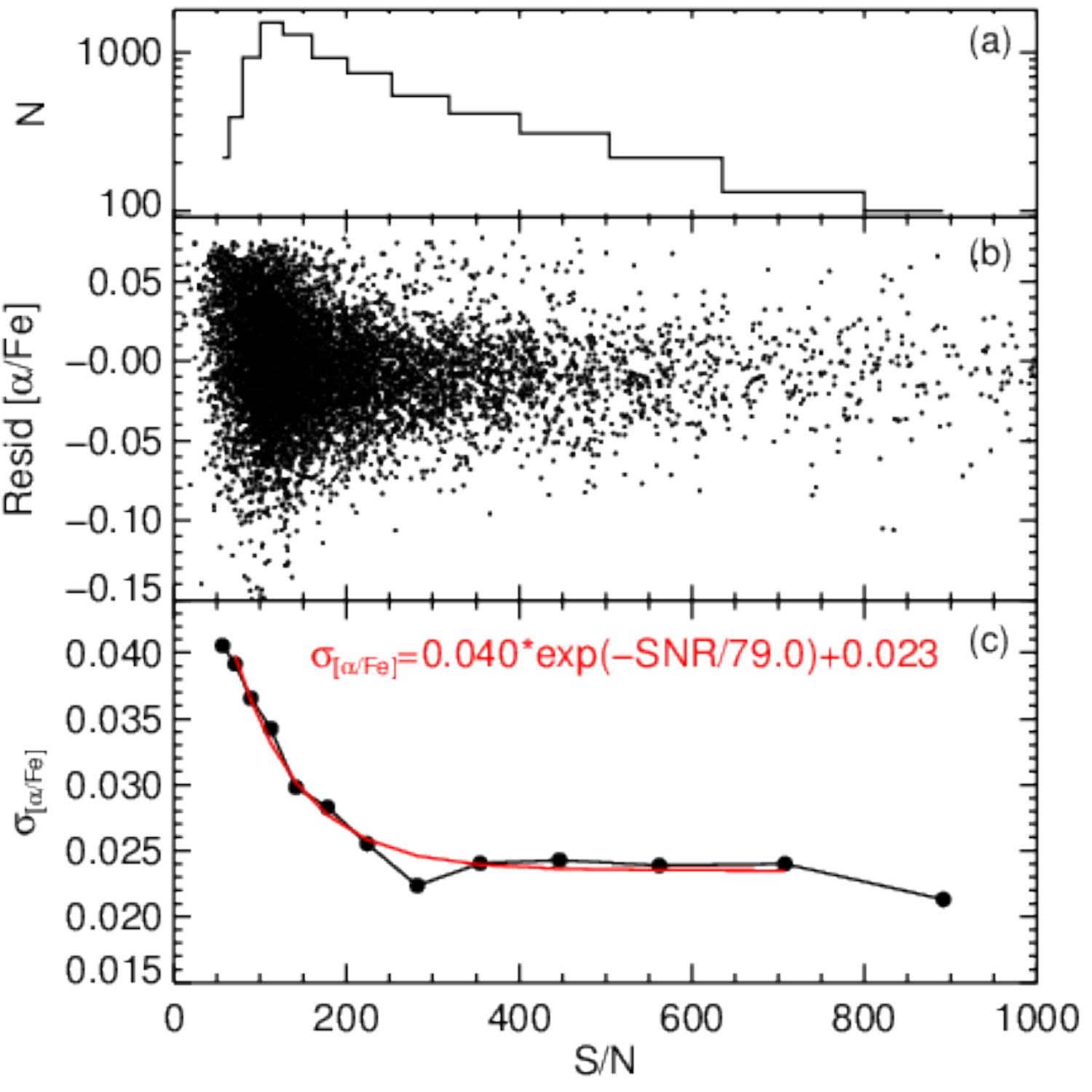}
\end{center}
\caption{The scatter in [$\alpha$/Fe] of the low-$\alpha$ sequence of stars ([$\alpha$/Fe]$<$$+$0.10; see Figure
\ref{fig_alphametals_all}) with $S/N$ after subtraction of the trend of [$\alpha$/Fe] with metallicity
(i.e., the ``banana'' shape).  (a) Histogram of low-$\alpha$ stars in each (logarithmic) bin of $S/N$. (b) The residual
[$\alpha$/Fe], after subtraction of the low-$\alpha$ [$\alpha$/Fe] trend with [Fe/H], versus $S/N$.
(c) $\sigma_{\rm [\alpha/Fe]}$, the robust standard deviation
of the residual [$\alpha$/Fe] in logarithmic bins of $S/N$, versus
$S/N$ showing an exponential decline with $S/N$.  The scatter reaches a plateau of 0.023 dex at $S/N$$\gtrsim$300.
The red line shows an exponential fit to the $\sigma_{\rm [\alpha/Fe]}$ values.}
\label{fig_sigalphasnr}
\end{figure}

The relationships between various elemental-abundance groups and between
abundances and age hold important clues about the evolution of the
various stellar populations constituting the MW disk. High-resolution
spectroscopic surveys, in combination with astrometry from Gaia
\citep{deBruijne2012} and high-precision asteroseismology data, will
allow these relations to be investigated over a much larger volume of
the disk than the local solar neighborhood, which has been the focus of
past surveys.  \citet{Anders14} studied the [$\alpha$/Fe] vs.\ [Fe/H]
distribution of red giants in three radial bins using the first year of
APOGEE data.
In this paper we go beyond this analysis by employing a sample of $\approx10,000$ red
clump stars with accurate distances from the first two years of the APOGEE survey to
investigate in detail
the relation between [$\alpha$/Fe] and [Fe/H] over a large part of the MW disk
using a large, statistical sample of stars spanning a wide range of ages
and a proper accounting for the targeting selection effects.
This unique sample allows tracing of
the locally-observed high- and low-$\alpha$ sequences toward and away
from the Galactic center, and to multiple kiloparsecs above the plane. 

This paper is organized as follows. In Section \ref{sec:red} we
discuss the APOGEE observations and data reduction, and we describe the
sample of red clump (RC) stars and biases pertaining to its selection in
Section \ref{sec:sample}.  Our results are presented in Section
\ref{sec:results}.  Chemical evolution models are discussed in Section
\ref{sec:gcemodel}, and the significance of our results is presented in Section
\ref{sec:discussion}.

In this study, we use Galactocentric rectangular coordinates (X,Y,Z) and
left-handed Galactocentric cylinderical coordinates (R,Z,$\phi$), assuming that the
Sun is 25 pc above the midplane and 8 kpc from the Galactic center,
as in the APOGEE--RC catalog paper \citep{Bovy14}.

\begin{figure}[ht!]
\begin{center}
\includegraphics[trim=0mm 8mm 10mm 8mm,clip,angle=0,scale=0.42]{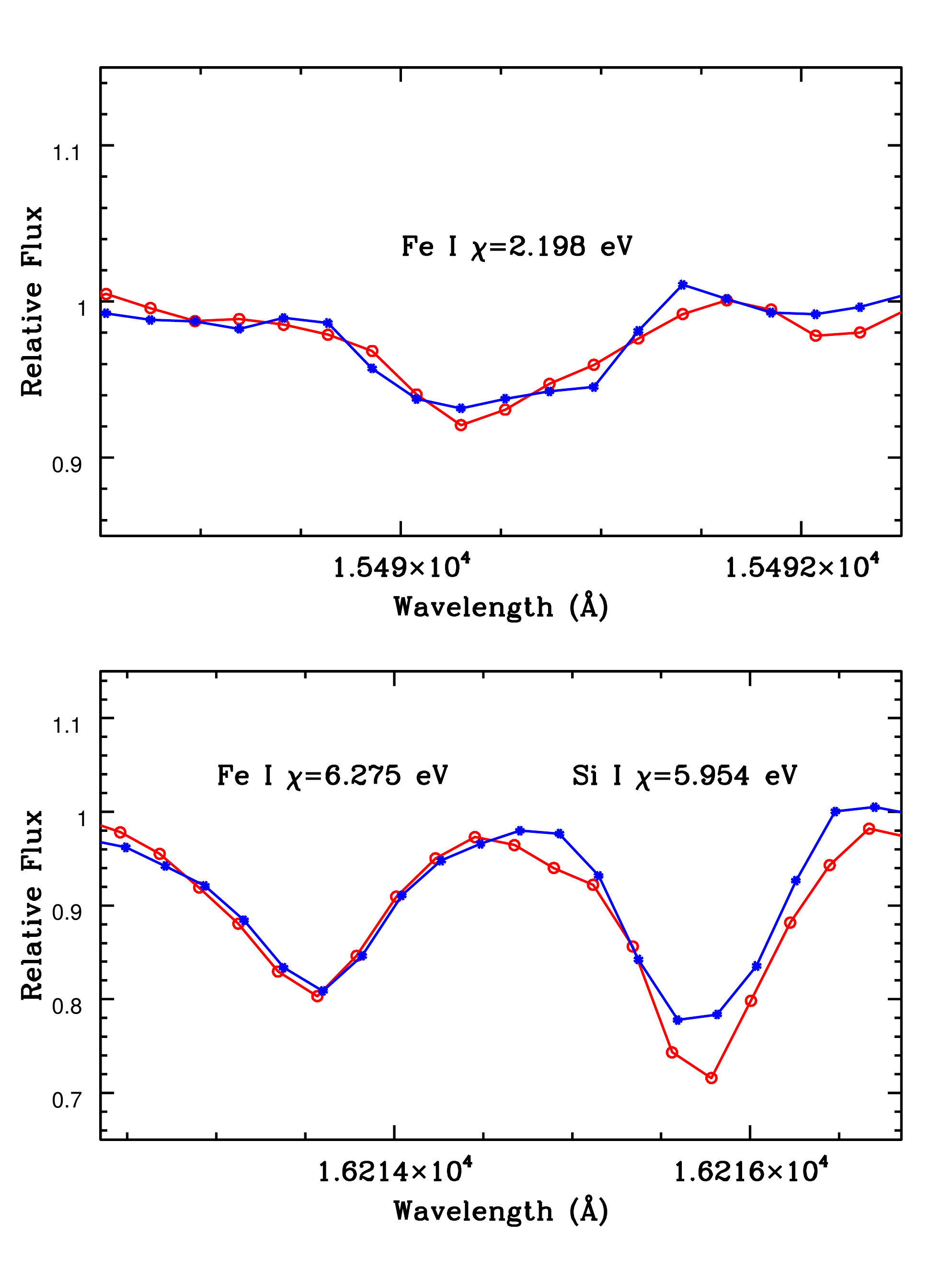}
\end{center}
\caption{Selected spectral lines from Fe~I and Si~I (an $\alpha$-element) in the
two red clump stars with different [$\alpha$/Fe] values used in the manual analysis.
In both 2M15152520+0102019 (red, $\alpha$-enhanced) and 2M06054047+2708560 (blue, solar-$\alpha$)
the two Fe~I lines have identical strengths, while the Si~I
line is significantly stronger in 2M06054047+2708560.  Two Fe~I lines are shown,
with one being low excitation and the other having a high excitation energy; the
fact that both lines have the same strength in both stars indicates that (1) their
effective temperatures are very similar and (2) they have nearly the same Fe abundance.
The stronger Si~I line in 2M15152520+0102019 shows that it is $\alpha$-enhanced
relative to the near-same Fe-abundance red clump star 2M06054047+2708560.}
\label{fig_manualanalysis}
\end{figure}

\begin{figure*}[ht!]
\begin{center}
\includegraphics[angle=0,scale=0.55]{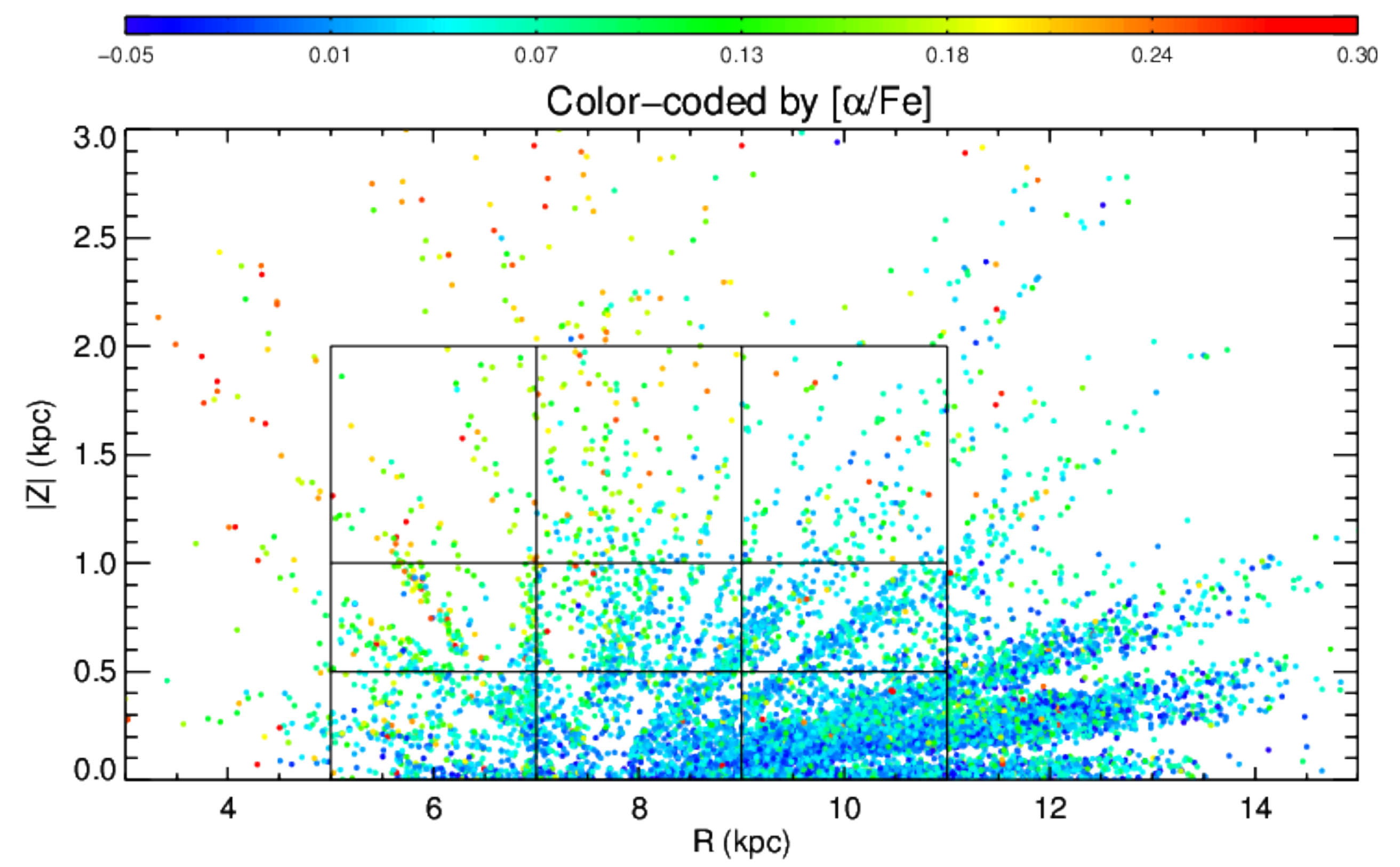}
\end{center}
\caption{$|Z|$ vs.\ $R$ distribution of APOGEE RC stars color-coded by [$\alpha$/Fe].  The nine $R/Z$ boxes used in
Figure \ref{fig_alphametals_rzbins} are shown.}
\label{fig_spatial_alpha}
\end{figure*}

\section{Observations And Data Reduction}
\label{sec:red}

We use the SDSS-III/APOGEE year 1 and 2 data for our analysis.  The APOGEE survey is
described in \citet{Eisenstein11} and Majewski et al.\ (2014, in preparation), and the instrument
in \citet{Wilson10,Wilson12}.  The data reduction is briefly described in \citet{Nidever12}, and will
be described in more detail in the near future (D.\ Nidever et al.\ 2014, in preparation).
Stellar parameters are derived for each star using a $\chi^2$ optimization algorithm with a large,
custom-built library of synthetic spectra \citep[][A.\ E.\  Garc{\'{\i}}a P{\'e}rez et al.\ 2014, in preparation]{Allende14}.
Pipeline versions similar to those used for Data Release 10 \citep{Ahn13} were also used
for the year 1 and 2 data that are the basis for this analysis\footnote{See \url{http://www.sdss3.org/dr10/irspec/}}.
The APOGEE pipelines produce reduced spectra, accurate radial velocities ($\sigma_{\rm RV}$$\sim$$0.1$ \kms for most
stars), and stellar parameters
that have been calibrated using globular clusters and other ``standards''
\citep{Meszaros13}.  
In the APOGEE synthetic spectral library the $\alpha$ elements O, Mg, Si, S, Ca, and Ti
are varied together in the $\alpha$ dimension of the 6-dimension grid.  Although the ASPCAP derived
[$\alpha$/Fe] abundance from the best-fit $\chi^2$ solution can often be thought of as the ``mean'' abundance
of these $\alpha$ elements, we find that in the limited \teff and \logg range of the RC sample the elements
O, Mg and Si are dominant.

The $H$-band has only recently been explored  for high-resolution spectroscopy
of late-type stars, and, therefore, detailed line formation studies in non Local Thermodynamical Equilibrium (LTE)
for this window and these stars are not available.
Departures from LTE are always a concern
in an analysis based on classical model atmospheres and LTE line
formation (i.e., assuming the source function is equal to the Planck
function). Specific calculations for each ion of interest
are needed to solve this issue. However, departures from LTE tend to be
reduced when the radiation field is weak (e.g., atmospheres of cool stars) and
the density is high (i.e., dwarf stars). In our case we deal with
cool stars with low gravities, and therefore some caution is advised.

We should note that working in the $H$-band, right
at the intersection where H$^{-}$ bound-free opacity yields
to H$^{-}$ free-free opacity at about 1.6 $\mu$m, the total opacity
reaches a minimum for cool stars.  This causes the continuum
to form in deeper atmospheric layers, and so do absorption lines,
which tend to be weak and in general form not too far from
the layers where the continuum forms. The higher density in those
layers, compared to those the optical spectrum is sensitive to,
favors the assumption of LTE.   At low metallicity, the lack of free
electrons makes it hard to couple matter with the radiation field.  Nonetheless,
lower metallicities bring higher pressure and
an increased role of collisions with hydrogen atoms. Unfortunately,
the collisional rates associated with inelastic hydrogen collisions
are set on firm theoretical basis only for a few of the lightest
ions \citep{Belyaev03,Belyaev10,Barklem12}.

While the APOGEE [$\alpha$/Fe] abundances
have not been thoroughly calibrated, an initial comparison of the [$\alpha$/Fe] abundances for stars
with literature values indicates that the APOGEE [$\alpha$/Fe] values are in line with expectations.
There are some peculiarities in the [$\alpha$/Fe] abundances for cool stars (\teff$\lesssim$4200 K),
that are generally difficult to analyze, but these effects are not apparent for warmer stars, including our RC
sample.  Indeed, for the stellar parameters typical of RC stars, the
differences between the best-fit APOGEE spectra and those obtained with
spherical MARCS models \citep{Gustafsson08} and Turbospectrum \citep[e.g.,][]{Plez12} are below 5\%.
The external uncertainties for [Fe/H] are $\sim$0.1 dex and $\sim$0.05 for [$\alpha$/Fe],
although these decrease with metallicity and are roughly $\sim$0.03--0.04 ([$\alpha$/Fe]) and $\sim$0.02--0.07
([Fe/H]) for the metallicities we consider here ($-0.9<$[Fe/H]$<+0.5$) \citep{Meszaros13}.  Figure
\ref{fig_sigalphasnr} shows the empirical scatter in [$\alpha$/Fe] for the low-$\alpha$ group
([$\alpha$/Fe]$<$0.10, $5\lesssim$$R$$\lesssim15$ kpc, $Z$$\lesssim$3 kpc, see Figure \ref{fig_alphametals_all} below)
around the [$\alpha$/Fe] trend with [Fe/H] (the ``banana'' shape) as a function of
$S/N$\footnote{$S/N$ per pixel in the combined ``apStar'' APOGEE spectrum with
roughly 3 pixels per resolution element.  The exact dispersion varies across the spectrum, but is on average
0.22\AA.}.  This scatter indicates the level of internal precision and 
decreases exponentially with $S/N$, reaching a plateau of $\sim$0.025 dex for $S/N$$\gtrsim$200;
$\sigma_{\rm [\alpha/Fe]}=0.040$ $\exp(-SNR/79.0)+0.023$.  The mean precision
of all RC stars based on the exponential fit is $\sim$0.027.
It is likely that the plateau in $\sigma_{\rm [\alpha/Fe]}$ at high $S/N$ is due to systematic effects in the spectra or
abundances, and indicates that the astrophysical scatter of stars in this sequence is below $\sim$0.02 dex.

\begin{figure*}[ht!]
\begin{center}
$\begin{array}{cc}
\includegraphics[angle=0,scale=0.40]{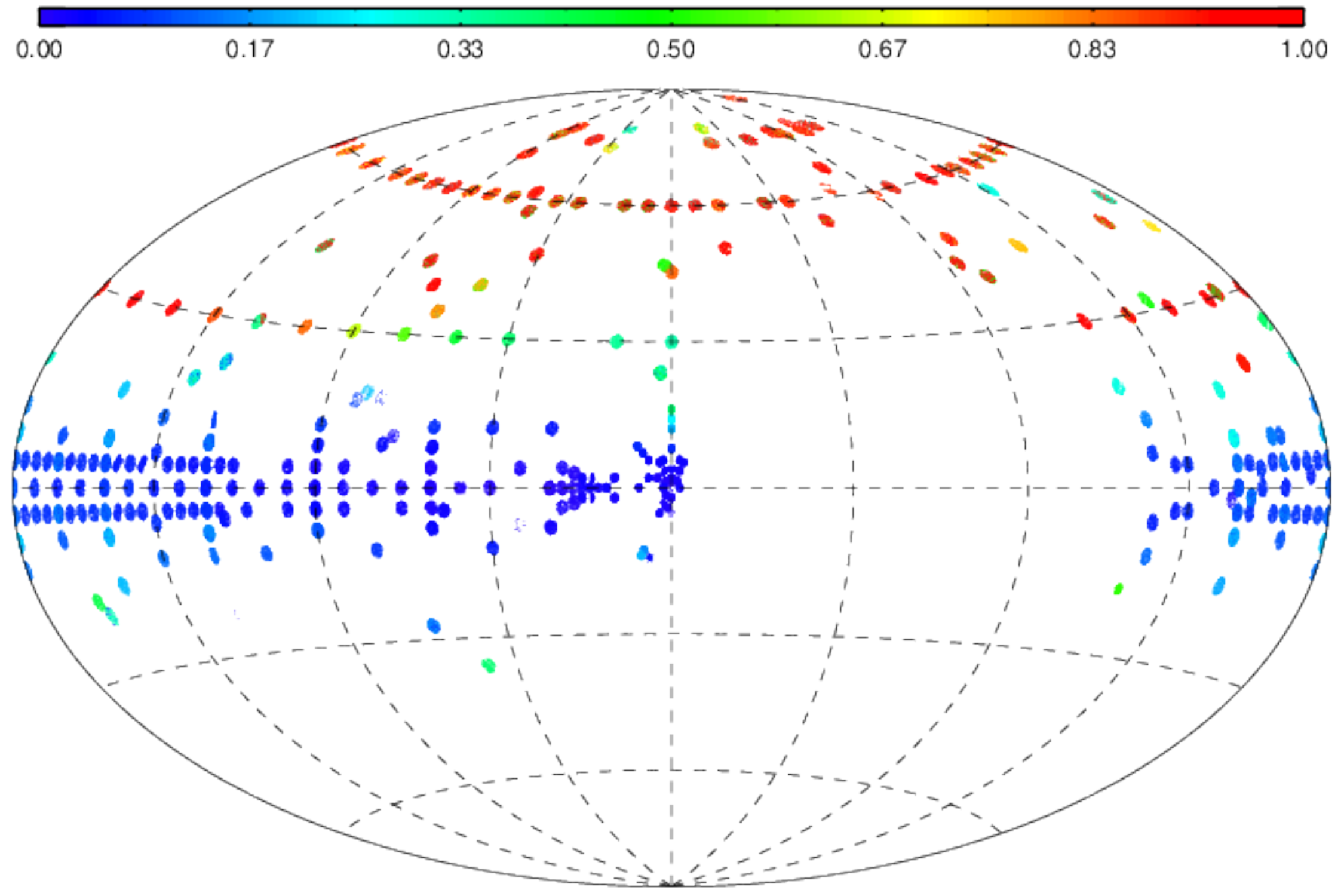}
\includegraphics[angle=0,scale=0.40]{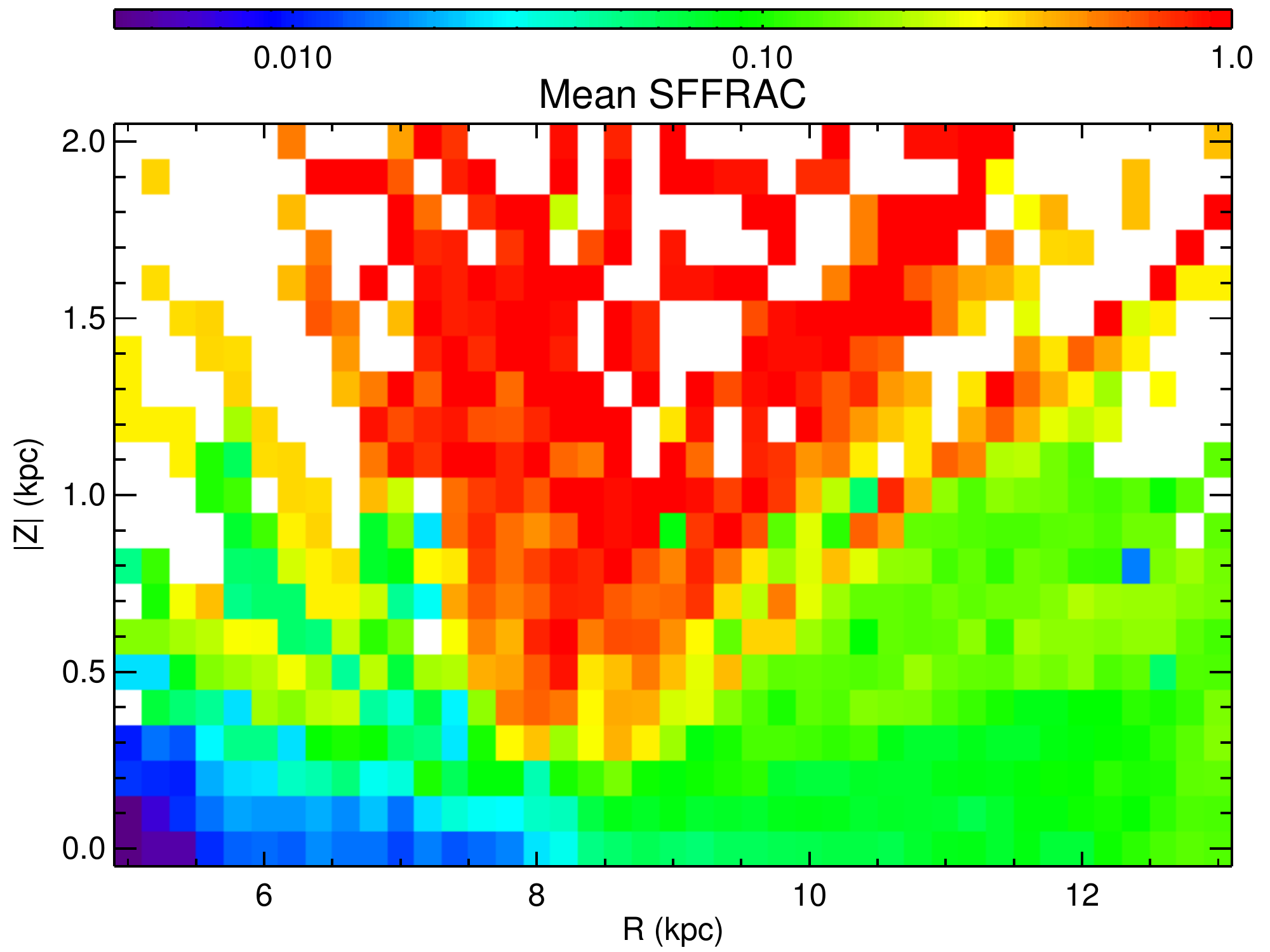}
\end{array}$
\end{center}
\caption{(left) The selection function fraction (SFFRAC) for all APOGEE stars in the statistical
sample.  (right) The mean selection function in the $R-|Z|$ plane for the RC stars (logarithmic scale).}
\label{fig_selfunc_targ}
\end{figure*}

\begin{figure}[ht!]
\begin{center}
\includegraphics[angle=0,scale=0.40]{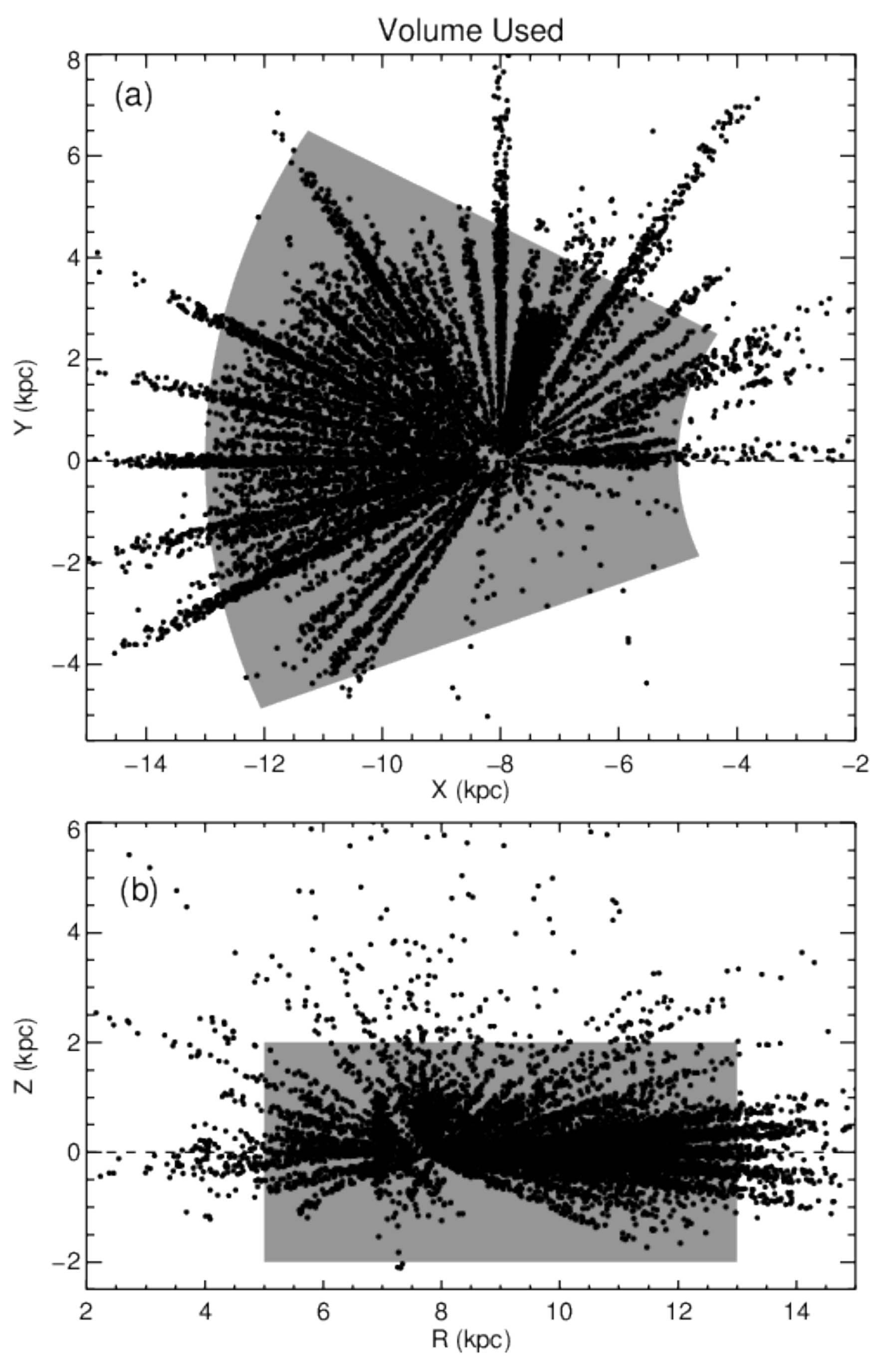}
\end{center}
\caption{The portion of the Galaxy explored by the APOGEE RC stars (solid dots); X--Y in (a) and $R$--$Z$ in (b).
The gray shading indicates the region for which the volume correction is calculated in Section \ref{sec:sample}: 
5 $\leq$ $R$ $\leq$ 13 kpc, $-$2 $\leq$ $Z$ $\leq$ $+$2 kpc, and $-$22\dgr $\leq$ $\phi$ $\leq$ $+$30\degr.}
\label{fig_selfunc_volcorr_volused}
\end{figure}

\begin{figure}[ht!]
\begin{center}
\includegraphics[angle=0,scale=0.40]{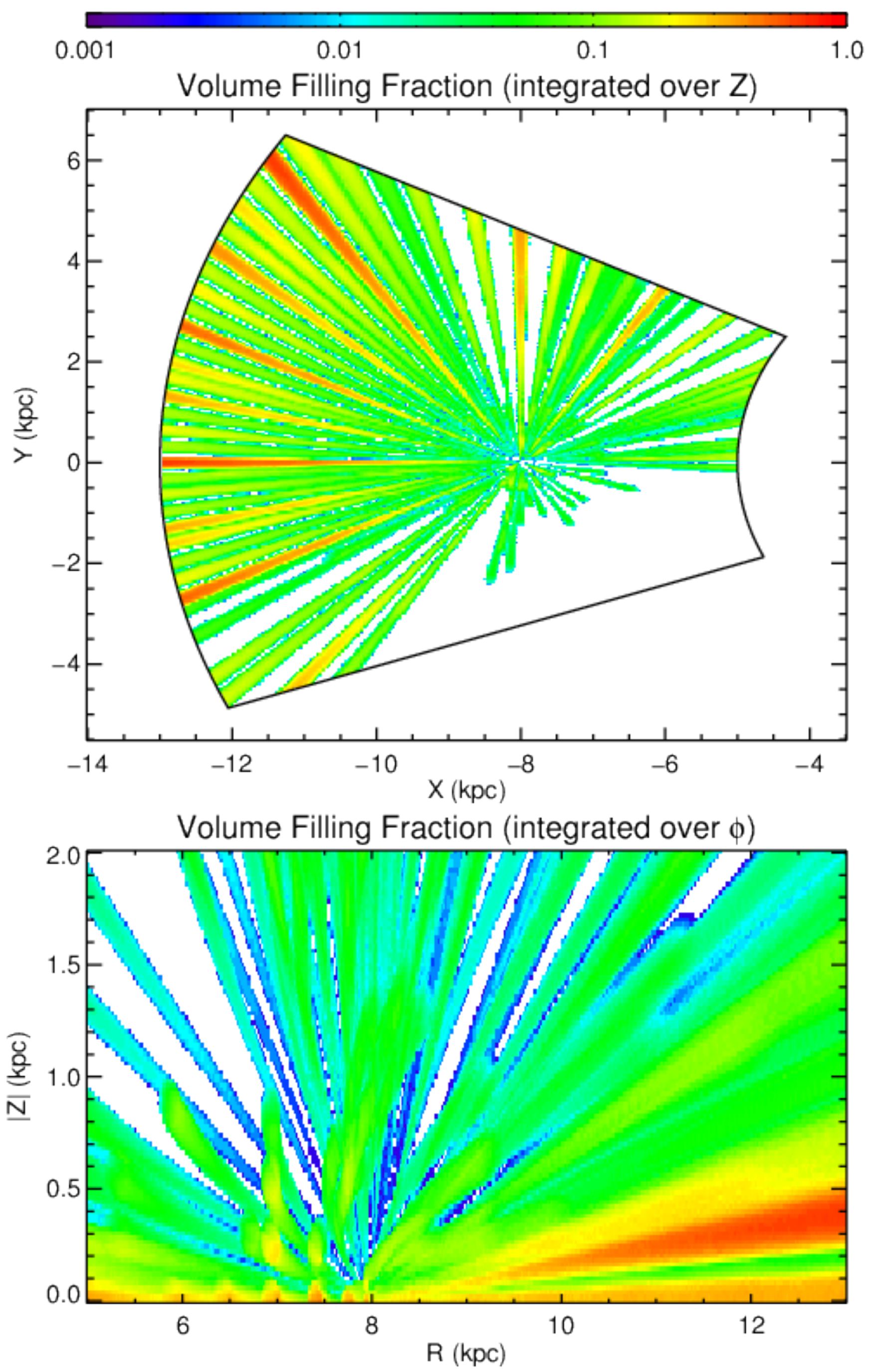}
\end{center}
\caption{The volume filling fraction in the (a) X-Y, and (b) $R-|Z|$ planes (logarithmic scales).}
\label{fig_selfunc_volcorr_fracrzxy}
\end{figure}

\subsection{Manual Consistency Check}

As a straightforward test of the ability of ASPCAP to differentiate stars
having different [$\alpha$/Fe] values, two RC stars with $S/N$$\approx$100 were chosen to be
subjected to a manual abundance analysis: 2M15152520+0102019 and 2M06054047+2708560.
These two stars were found to have very similar stellar parameters by
ASPCAP (with typical RC values), but differing [$\alpha$/Fe]
values.  In the case of 2M15152520+0102019, ASPCAP finds \teffe=
4891K, \logg= 2.65, [Fe/H]= $-$0.46, and [$\alpha$/Fe]= $+$0.33.  The other
star, 2M06054047+2708560, has ASPCAP values of \teffe= 4895K,
\logg= 2.64, [Fe/H]= $-$0.45, and [$\alpha$/Fe]= $-$0.03.

\begin{figure}[ht!]
\begin{center}
\includegraphics[angle=0,scale=0.42]{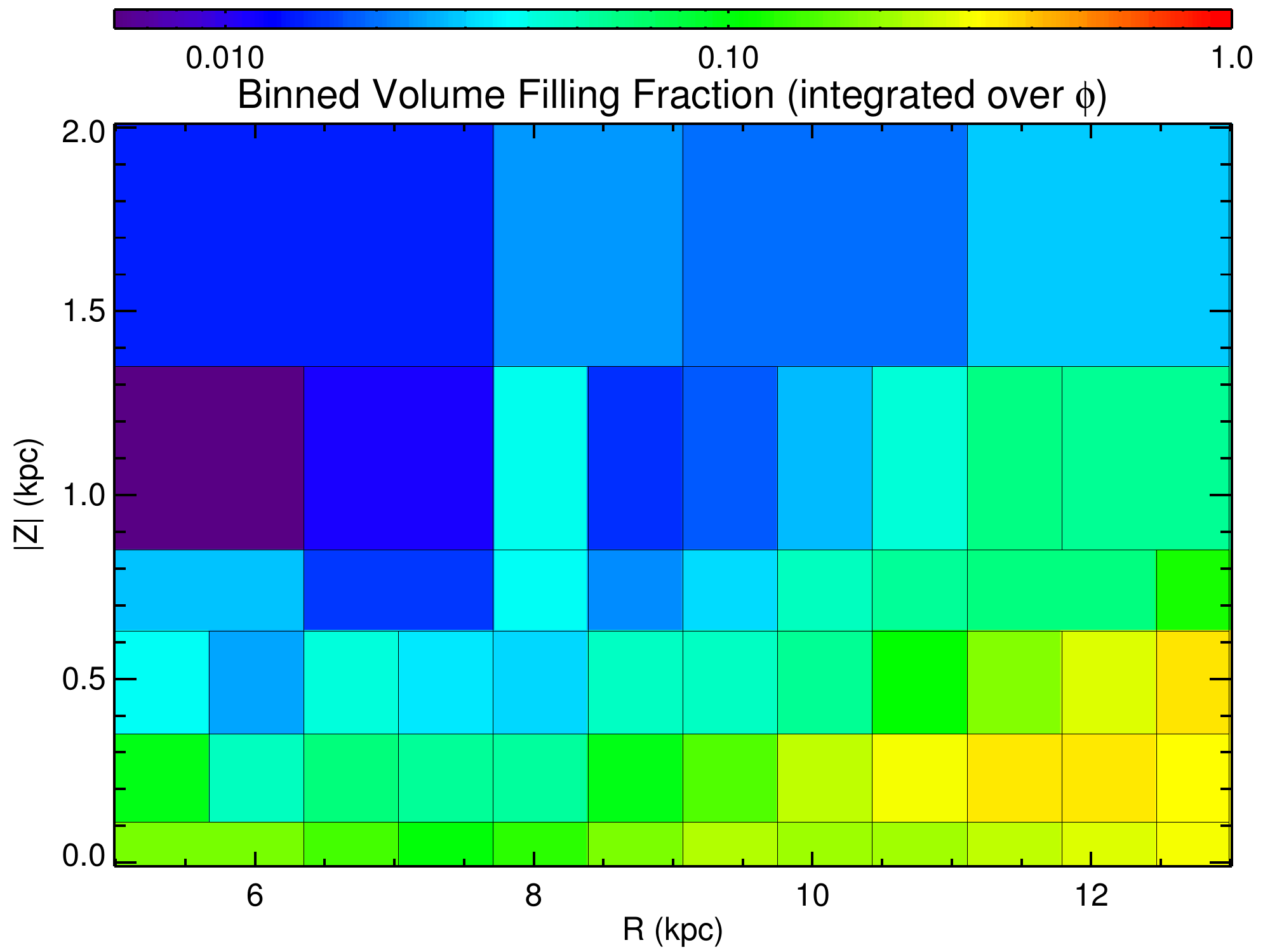}
\end{center}
\caption{The binned volume filling fraction in the $R-|Z|$ plane (logarithmic scale).}
\label{fig_selfunc_volcorr_fracrz_bin}
\end{figure}

\begin{figure}[ht!]
\begin{center}
\includegraphics[angle=0,scale=0.42]{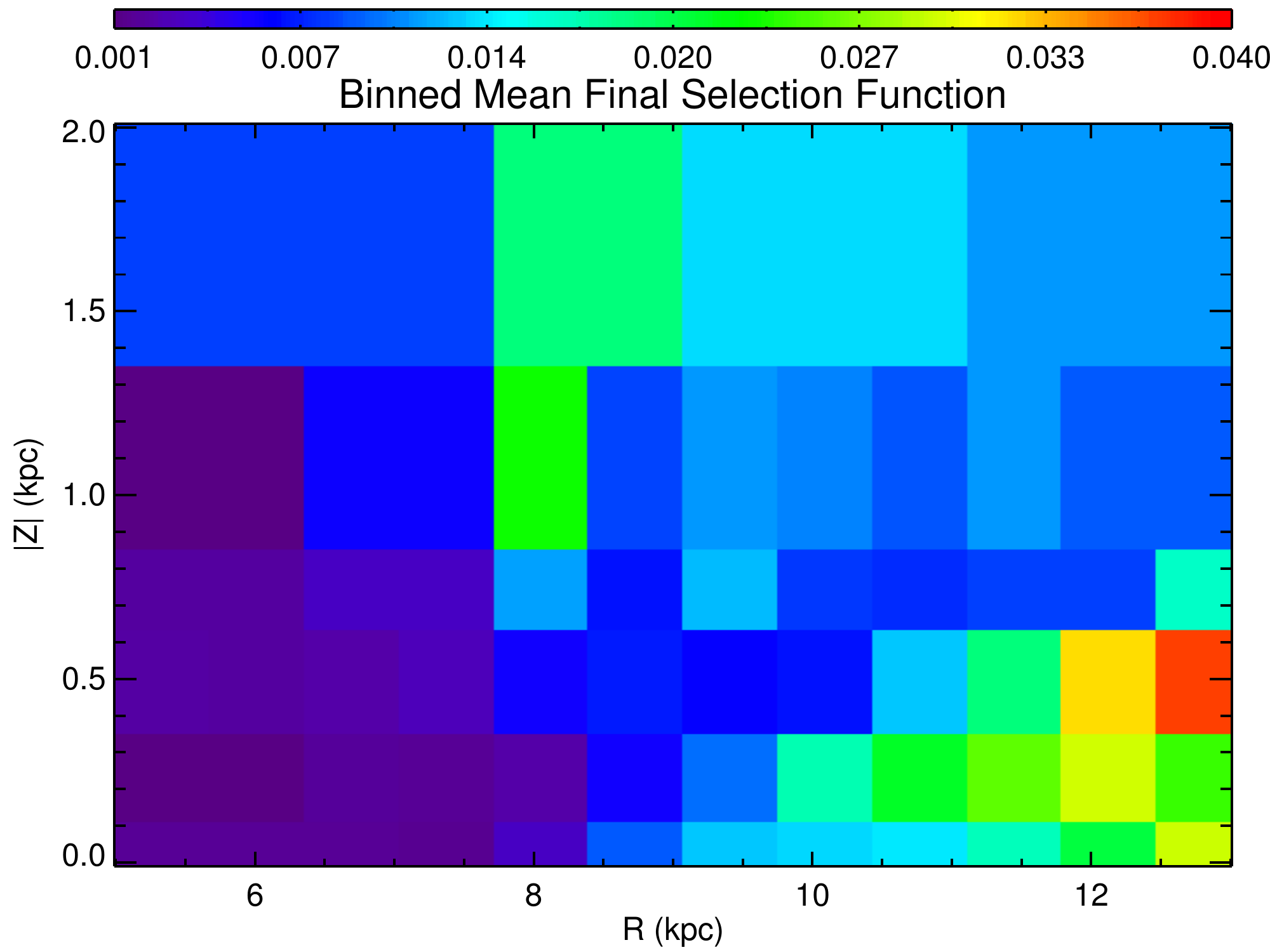}
\end{center}
\caption{The binned mean final selection function in the $R-|Z|$ plane (linear scale).}
\label{fig_selfunc_combcorr_sffracrz}
\end{figure}

Model atmospheres from the same grid (ATLAS9, scaled-solar abundance
models) as that used for ASPCAP were interpolated, and an LTE abundance
analysis using the spectrum synthesis code MOOG was carried out for the
elements iron, magnesium, silicon, calcium, and titantium.  The spectral
lines chosen to be synthesized were those listed in \citet{Smith13},
who explored spectral windows for various elements using the APOGEE 
wavelength range and APOGEE linelist in an analysis of field red giant
standard stars.
The elements Mg, Si, Ca, and Ti are taken to represent the
$\alpha$-elements.  As ASPCAP in its current form averages the abundances
of the $\alpha$-elements together to form a single [$\alpha$/Fe] index,
the same average is computed here.  Synthesis of the Fe~I, Mg~I, Si~I,
Ca~I, and Ti~I lines was carried out for both RC stars as done in
\citet{Smith13}, and the results for each element are now noted for each
star.  For 2M15152520+0102019 the abundances are: [Fe/H]= $-$0.44$\pm$0.09,
[Mg/H]= $-$0.24$\pm$0.15, [Si/H]= $-$0.26$\pm$0.07, [Ca/H]= $-$0.17$\pm$0.03,
and [Ti/H]= $-$0.15$\pm$0.02, leading to [$\alpha$/Fe]= $+$0.24$\pm$0.10.  For 
2M06054047+2708560 the abundances are: [Fe/H]= $-$0.48$\pm$0.11, [Mg/H]=
$-$0.36$\pm$0.18, [Si/H]= $-$0.44$\pm$0.07, [Ca/H]= $-$0.48$\pm$0.04, and
[Ti/H]= $-$0.57$\pm$0.04, which then yields [$\alpha$/Fe]= $+$0.02$\pm$0.12.

\begin{figure*}[ht!]
\begin{center}
\includegraphics[angle=0,scale=0.60]{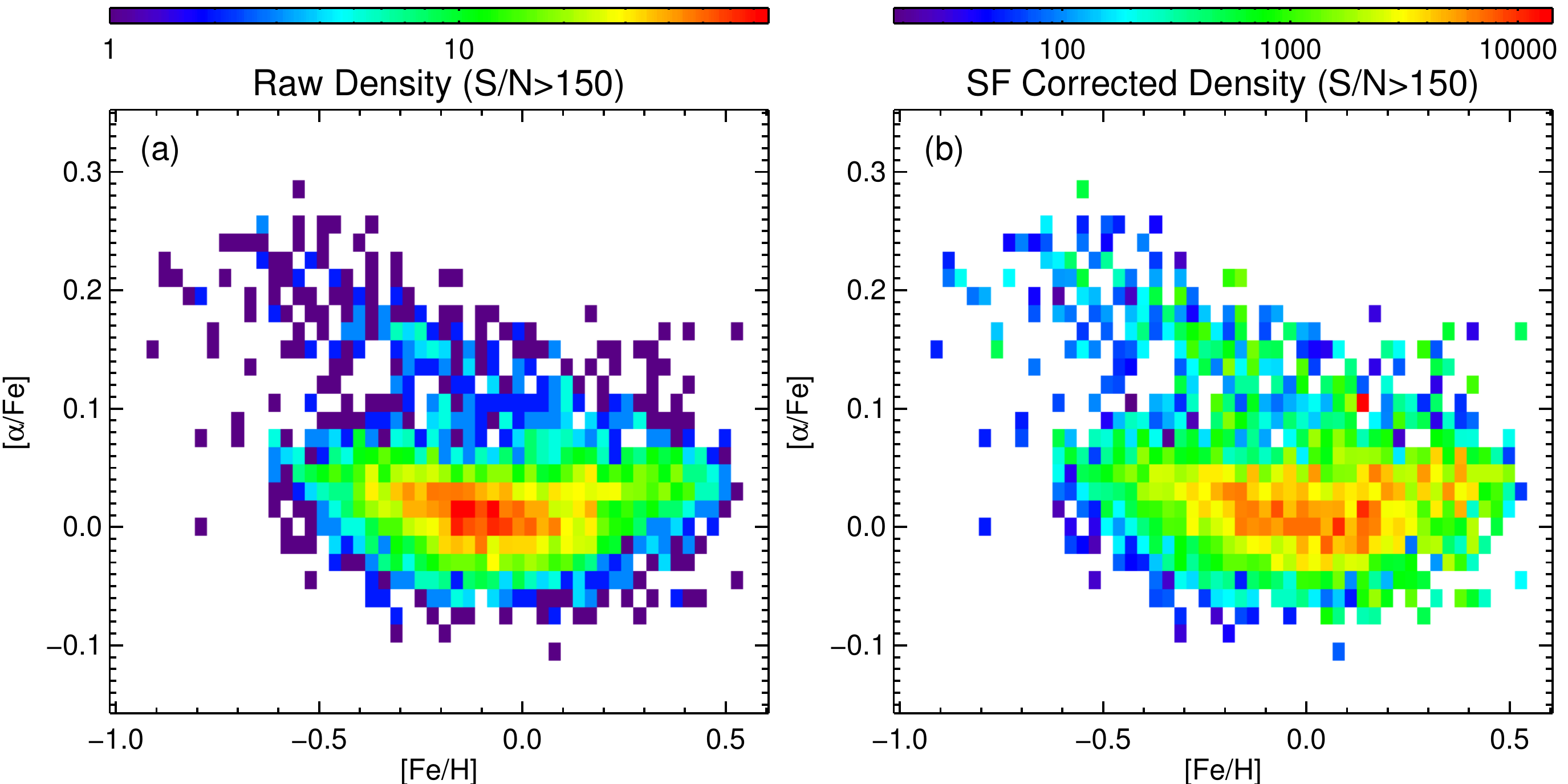}
\end{center}
\caption{The effect of the selection function on the abundance patterns in the [$\alpha$/Fe] vs.~[Fe/H]
plane. (a) The raw density of RC stars, and (b) the selection function corrected density of RC stars.}
\label{fig_selfunc_alphametals}
\end{figure*}

The manual analysis confirms the ASPCAP result that both RC stars
have similar Fe abundances, while 2M15152520+0102019 is $\alpha$-enhanced
and 2M06054047+2708560 has a [$\alpha$/Fe] value that is solar.  The differences   
between these two stars is illustrated visually in Figure \ref{fig_manualanalysis}, where
selected Fe~I and Si~I lines are over-plotted for the two stars.  Both stars have
near-identical Fe I lines (both low-excitation and high-excitation energies),
with 2M15152520+0102019 exhibiting a significantly stronger Si I line, resulting
in a larger Si abundance in this star.  Similar differences are found in the
Mg~I, Ca~I, and Ti~I lines, demonstrating that 2M15152520+0102019 has an elevated
value of [$\alpha$/Fe]. 
The ASPCAP abundances are very similar to those provided by the manual
analysis: for 2M15152520+0102019 $\Delta$[Fe/H](ASPCAP$-$Manual) = $-$0.02
dex, while 2M06054047+2708560 has $\Delta$[Fe/H](ASPCAP$-$Manual) = $+$0.03
dex.  In the case of the ratios of [$\alpha$/Fe], 2M15152520+0102019 has
$\Delta$[$\alpha$/Fe](ASPCAP$-$Manual) = $+$0.09 dex and 2M06054047+2708560
has $\Delta$[$\alpha$/Fe](ASPCAP$-$Manual) = $-$0.05 dex.  In all abundances
and their ratios, the differences are $\lesssim$ 0.1 dex (i.e., less than the measurement
uncertainties of the manual analysis), and indicate that
ASPCAP is deriving reliable stellar parameters, metallicities,
and mean $\alpha$-to-Fe ratios for RC stars.
A more detailed analysis of ASPCAP abundance uncertainties will be presented in
A. E. Garc{\'{\i}}a P{\'e}rez et al.\ (2014, in preparation).

\section{Sample and Selection Effects}
\label{sec:sample}

For this analysis we use the He-core burning red clump stars observed by APOGEE.  RC stars are
in general $\sim$3--4 times more numerous than stars on the upper red giant branch (RGB, brighter than the RC).
Due to APOGEE's simple targeting color cuts, ($J$-$K_{\rm s}$)$_0$$>$0.5, many RC stars have been observed by
APOGEE \citep{Zasowski13}.  An advantage conferred by use of RC stars for a spatial survey
of stars is that their absolute magnitudes vary little with age and metallicity
\citep{Girardi02,Groenewegen08}, i.e., they are ``standard candles'', and their distances can be readily
and reliably inferred.  The RC stars also have the added benefit of being warm enough that their APOGEE [$\alpha$/Fe]
abundances are reliable and don't suffer from systematic effects currently seen in the APOGEE abundances of
cool giants \citep{Anders14,Hayden14a}.

The RC sample selection is described in detail in our companion paper on the APOGEE--RC catalog \citep{Bovy14}.
In brief, we select RC stars using simple selections in \logg as a function of \teffe, and in dereddened
color as a function of metallicity.  A comparison with APOGEE stars that have accurate surface-gravity measurements
from {\it Kepler} astroseismology (M. Pinsonneault et al.\ 2014, in preparation) demonstrates that our RC sample suffers
$\lesssim$7\% ``contamination''
from RGB stars.  The RC $K_{\rm s}$-band magnitudes are extinction-corrected using 2MASS+$Spitzer$/WISE photometry
and the RJCE method \citep{Majewski11}.  PARSEC models \citep{Bressan12} are employed to determine the absolute magnitude of
the RC stars using ([Fe/H],[J-K$_{\rm s}$]$_0$).  The derived RC
distances are accurate to $\sim$5\%.  The majority of RC stars are within $\sim$4--5 kpc from the sun,
but some extend out to $\sim$10 kpc.  The final catalog has 10,341 RC stars and their distribution in the
$|$Z$|$--$R$ plane is shown in Figure \ref{fig_spatial_alpha}.

\begin{figure*}[ht!]
\begin{center}
$\begin{array}{cc}
\includegraphics[angle=0,scale=0.40]{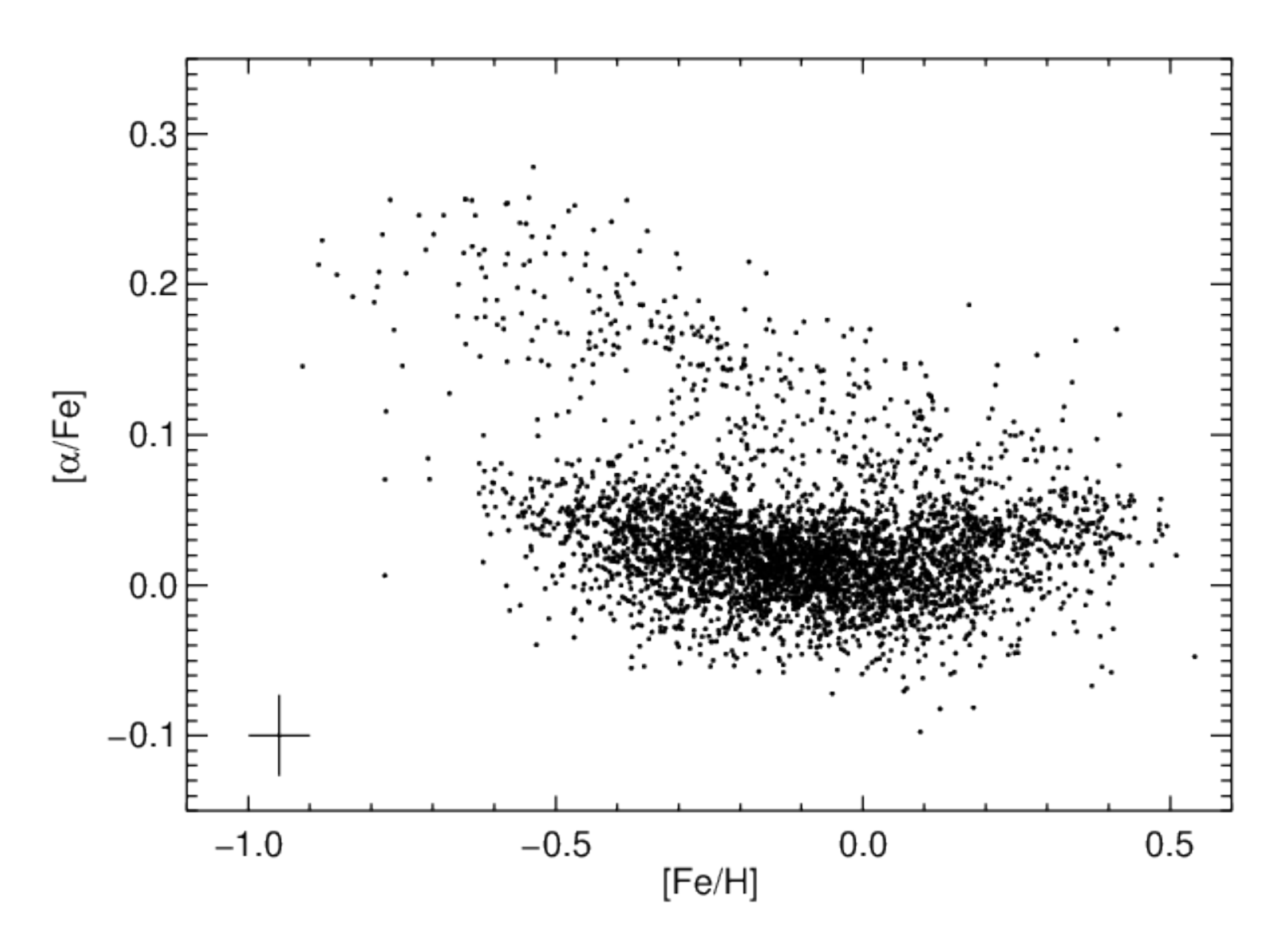}
\includegraphics[angle=0,scale=0.40]{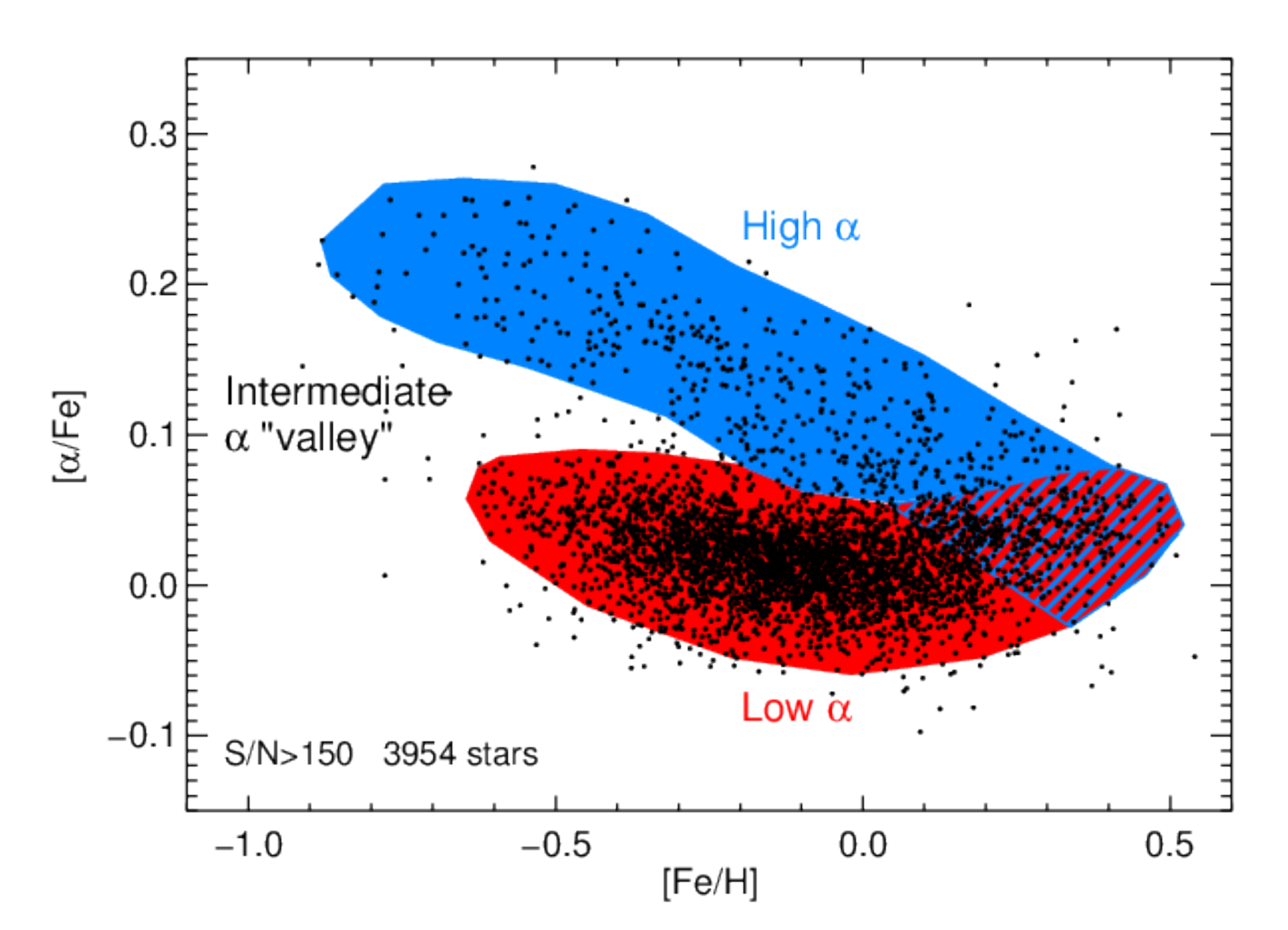}
\end{array}$
\end{center}
\caption{(left) The [$\alpha$/Fe] vs.\ [Fe/H] diagram for our sample of APOGEE RC stars
with $S/N$$>$150 (3954 stars).
The bimodality in [$\alpha$/Fe] at low [Fe/H] is clearly visible and extends over $\sim$0.6 dex in metallicity.
The error bar in the lower left corner shows representative precisions of 0.05 dex in [Fe/H] and
0.027 dex in [$\alpha$/Fe].
(right) A schematic of our RC [$\alpha$/Fe] vs.\ [Fe/H] diagram showing the main features: the
low-$\alpha$ group (red), high-$\alpha$ group (blue), and intermediate-$\alpha$ ``valley''.
The hashed red/blue shows the overlap region between the low- and high-$\alpha$ stars.
The black dots are RC stars with $S/N$$>$150.}
\label{fig_alphametals_all}
\end{figure*}

\begin{figure*}[ht!]
\begin{center}
\includegraphics[angle=0,scale=0.58]{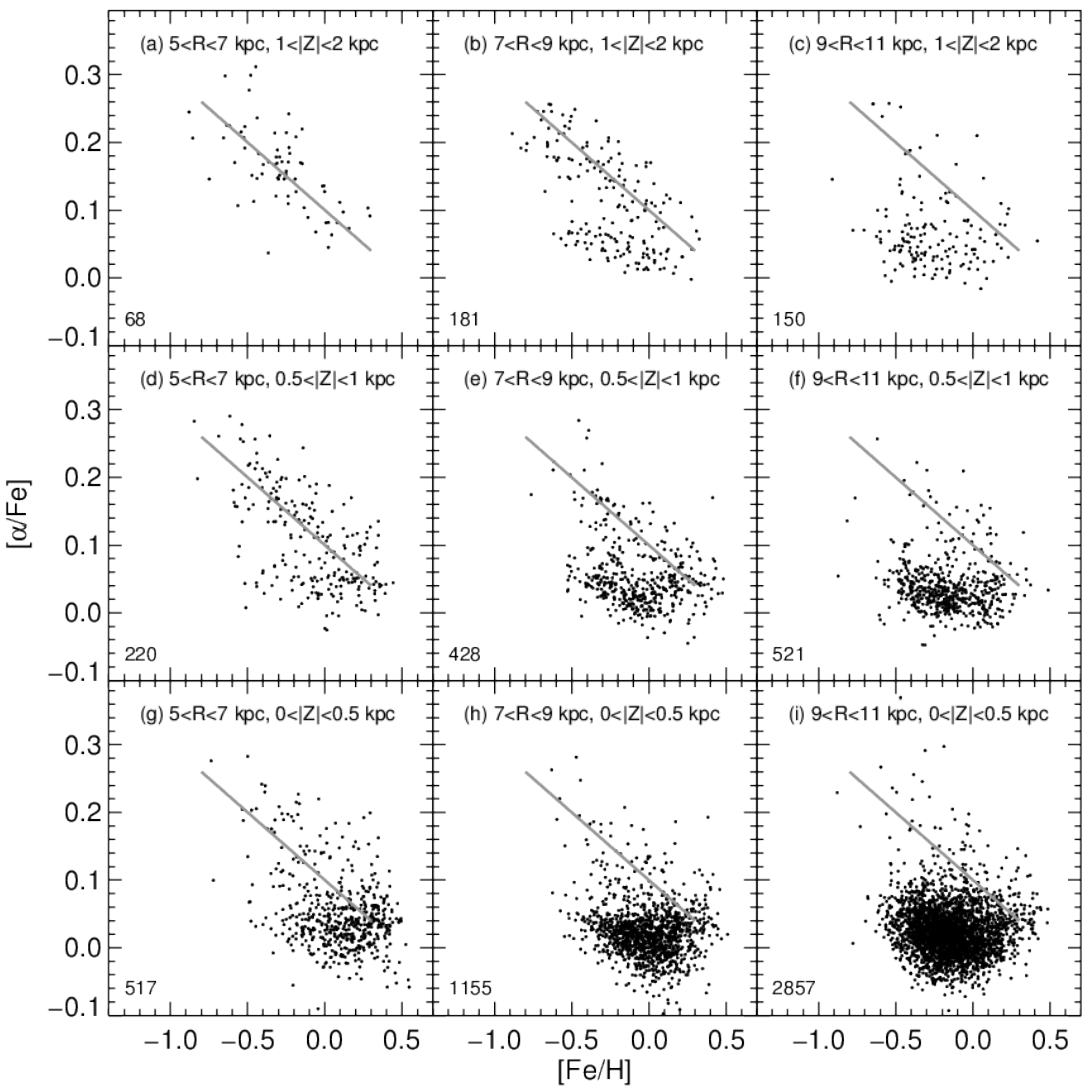}
\end{center}
\caption{The [$\alpha$/Fe] vs.\ [Fe/H] diagram of our RC stars, similiar to Figure \ref{fig_alphametals_all},
but now separated into nine different Galactic $R$ and $Z$ zones.  A high-$\alpha$ sequence fiducial gray line,
[$\alpha$/Fe]=$-$0.2$\times$[Fe/H]$+$0.10,
is drawn in all the panels.  The number of stars in each spatial zone is shown in the bottom left corner.}
\label{fig_alphametals_rzbins}
\end{figure*}

The APOGEE targeting scheme is described in detail in \citet{Zasowski13}.
There are three main selection effects introduced by the targeting scheme used by APOGEE:
\begin{enumerate}
\item The color cut of ($J$-$K_{\rm s}$)$_0$$>$0.5.  This removes RC stars with [Fe/H]$<-0.9$, but only
  marginally affects our results because most MW disk stars are more metal-rich than this cutoff
(see Section 5 of \citealt{Bovy14}).
\item The magnitude groups or ``cohorts''.  APOGEE targets are selected in three magnitude
ranges: 7$<$$H$$<$12.2 (short), 12.2$<$$H$$<$12.8 (medium), and 12.8$<$$H$$<$13.3/13.8 (long) with 3/6/12--24
``visits''\footnote{Each ``visit'' is roughly one hour of integration time.} per group respectively
to attain net $S/N$=100 for stars in each group.  Each magnitude range is divided into three magnitude bins
with equal numbers of stars and then 1/3 of the stars allocated for that magnitude range are randomly drawn from 
each of the three bin (see \citealt{Zasowski13} for more details).  Because the brighter stars reach their $S/N$
threshold more quickly, multiple
groups of stars, or cohorts, are observed in that range over the many visits to that field.  Due to this
scheme, a larger number of brighter (closer) stars are observed than fainter (distant) stars.  In general, the 230
science fibers allocation by cohort is roughly 90/90/50, although this distribution varies with $l$ and $b$.
\item The placement of fields in the sky.  APOGEE uses a fairly uniform grid of fields in Galactic
($l$,$b$), as seen in \citet{Zasowski13} and the left panel of Figure \ref{fig_selfunc_targ}.
\end{enumerate}

The effects from (2) and (3) create spatial biases in the APOGEE sampling of the RC stars across the MW.
However, the selection bias is only in apparent magnitude, and, because RC stars are standard candles within
$\sim$0.05 mag, the bias is in RC distance and the
Galactic coordinates $R$ and $Z$.  If the abundances of stars vary with position in the Galaxy, which
has been shown by previous studies \citep[e.g.,][]{Cheng12a,Bovy12a,Schlesinger12,Anders14,Hayden14a,Boeche14}, this
positional bias can create a chemical bias.  If, however, the analysis is restricted to small positional zones,
or the abundances explicity plotted with their dependence on position, then the bias is effectively removed.

\begin{figure}[ht!]
\begin{center}
\includegraphics[angle=0,scale=0.53]{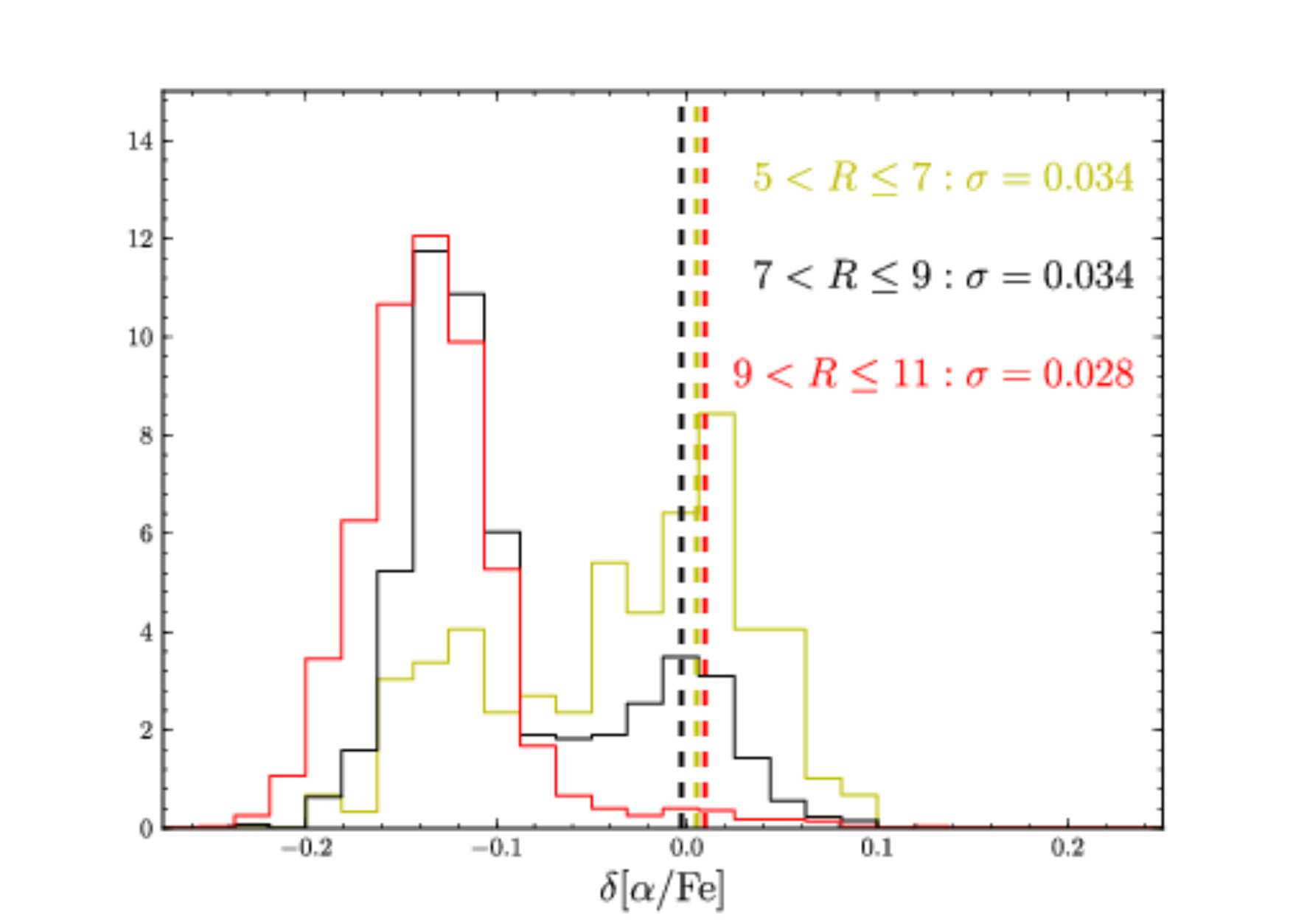}
\end{center}
\caption{The deviation of stars from the high-$\alpha$ trend line, $\delta$[$\alpha$/Fe] vs.\ [$\alpha$/Fe].
Stars with $S/N$$>$70, $|Z|$$\leq$3 kpc, $-0.6$$\leq$[Fe/H]$\leq$$-0.2$ are used in three radial bins.
The trend line is $-0.2\times$[Fe/H]$+0.10$.
}
\label{fig_deltaalphahist_rbins}
\end{figure}

We correct for the APOGEE selection effects in two steps: (1) the stars in our fields that could have been
observed but were not, and (2) the survey coverage or volume correction, i.e., stars in fields that we did not
observe.  The first step
is fairly straightforward, and is described in detail in $\S$4 of the RC catalog paper.  For every adopted
target selection cut we count the number of stars that passed the cut and compare this number to the actual number
of stars observed by APOGEE for this field; the ratio (N$_{\rm observed}$/N$_{\rm total}$) is the selection function
fraction (SFFRAC) for these APOGEE-observed stars.  In our disk and bulge fields, there is a simple
($J$-$K_{\rm s}$)$_0$$>$0.5 color cut and magnitude limits for the cohorts.  However, in the halo fields,
supplemental Washington M, T$_2$ and DDO51
\citep[W+D;][]{Majewski00} photometry was often used to pre-select giant stars \citep{Zasowski13}.
The target selection in
these fields included whether 2MASS-detected stars were identified as giants based on the W+D photometry and was
therefore more complex.  However, the method for calculating the selection effects remains the same.
The selection function fraction was computed for stars in all fields that were selected as part of the
APOGEE ``survey'' or ``statistical'' (as it is referred to in the RC catalog paper) sample (i.e., not {\it Kepler} fields,
ancillary targets, globular cluster members, etc.).
The left panel of Figure \ref{fig_selfunc_targ} shows the SFFRAC in the APOGEE fields explored here (year 1+2). 
The SFFRAC values are low for the low-latitude fields (especially in the inner galaxy) because there are far so many thin
disk stars that could be targeted according to the color and magnitude selection criteria.
The opposite is seen in the higher-latitude ($|$b$|$$>$16\degr), ``halo'' fields where
the values are high because the overall density of stars is low and all possible stars are observed.
The right panel of Figure \ref{fig_selfunc_targ} shows the mean SFFRAC of the RC stars in the $R-|Z|$ plane.  

\begin{figure}[ht!]
\begin{center}
\includegraphics[angle=0,scale=0.38]{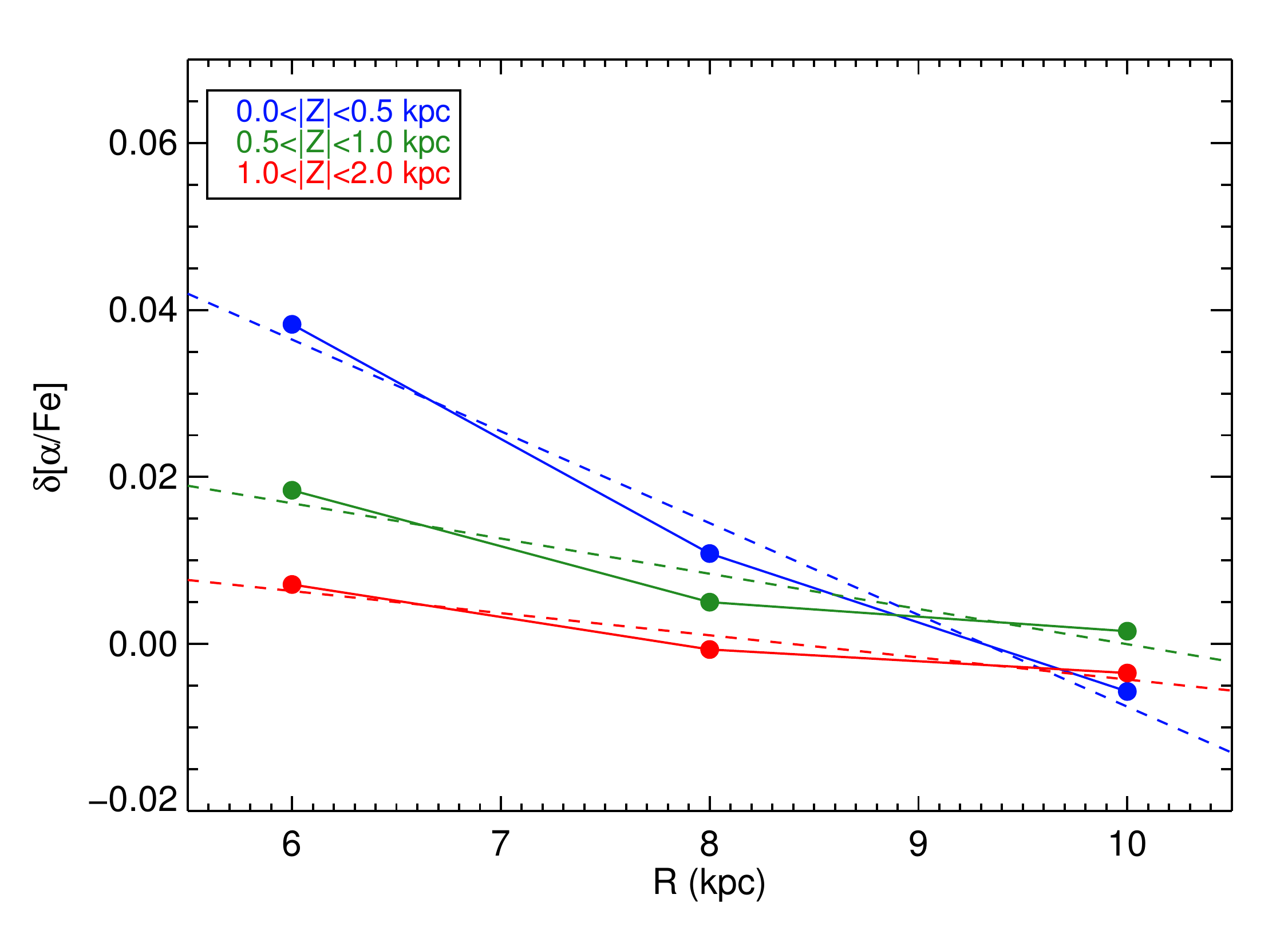}
\end{center}
\caption{The median $\delta$[$\alpha$/Fe] (deviation from the high-$\alpha$ trend line) 
for high-$\alpha$ stars ([$\alpha$/Fe]$>$0.09) separated for our nine $R/Z$ zones
(as seen in Figure \ref{fig_alphametals_rzbins}) versus $R$ with zones at the same $|Z|$
connected by lines (and color-coded).
Dashes lines show the best-fit linear trend of $\delta$[$\alpha$/Fe] with $R$ for the three zones
at the same $|Z|$.  There is a negative gradient of $\delta$[$\alpha$/Fe] with $R$ at any $|Z|$
with a slope that decreases with $|Z|$. }
\label{fig_afeshift_trends}
\end{figure}

The second step is to calculate the fraction of the Galactic volume that is explored with the APOGEE
RC stars we actually did probe as a function of $R-|Z|$.  We calculate a fine grid (steps of 20 pc in $R$/$Z$ and 0.2\dgr
in $\phi$) in Galactic cylindrical coordinates for the region explored by the APOGEE RC stars: 5 $\leq$ $R$ $\leq$ 13 kpc,
$-$2 $\leq$ $Z$ $\leq$ $+$2 kpc, and $-$22\dgr $\leq$ $\phi$ $\leq$ $+$30\dgr (gray region in Figure
\ref{fig_selfunc_volcorr_volused}).  We then flag every voxel (pixel in our 3D grid) that falls within the cone of one
of the APOGEE survey fields.
The volume filling fraction in the X-Y (integrated over $Z$) and
$R-|Z|$ (integrated over $\phi$) planes are shown in Figure \ref{fig_selfunc_volcorr_fracrzxy}.  These values show
the opposite trends of SFFRAC.  APOGEE covers a large fraction of the volume in the midplane but a significantly
lower fraction at high latitude.  However, there are hardly any holes in the $R-|Z|$ plane except for a few
gaps at high-latitude in the inner galaxy.  To correct for these missed regions we bin the volume fractions in the 
$R-|Z|$ plane with zones of $\sim$45 stars or more (Figure \ref{fig_selfunc_volcorr_fracrz_bin}).  All RC stars
in the above-mentioned volume are given volume correction fractions based on the binned image.

The product of the two selection functions, SFFRAC and the volume filling fraction, is taken to produce the
final selection function as seen in Figure \ref{fig_selfunc_combcorr_sffracrz} (on a linear scale).  These
values show some variations in the $R-|Z|$ plane but they are not large and vary smoothly with
position.  To assess the impact of the APOGEE selection function on the abundance patterns in the [$\alpha$/Fe]
vs.~[Fe/H] plane, we compare the raw density to the selection function corrected density (each star weighted
by the inverse of the final selection function) of RC stars seen in Figure \ref{fig_selfunc_alphametals}.
The selection function correction changes the relative numbers of stars in each abundance group but the overall
{\em pattern} of abundances does not change.  This effect is especially clear when comparing to the unbinned plot
in the left panel of Figure \ref{fig_alphametals_all}.  Because in this paper we are focused on the patterns
in the abundance plane and information is lost by binning, we use the raw scatter plots for the rest
of the paper.

\section{Results}
\label{sec:results}

\subsection{Full Sample}

Figure \ref{fig_alphametals_all} displays the [$\alpha$/Fe] vs.\ [Fe/H] diagram for all of our
APOGEE RC stars with $S/N$$>$150.  The $\alpha$-bimodality is clearly visible for $-0.9<$[Fe/H]$<-0.2$ dex.
Our $\alpha$-element abundance distribution shows some differences compared to the medium-resolution SEGUE G-dwarf sample
of \citet{Lee11} and \citet{Bovy12b}, especially in the distinct bimodality at the metal-poor end.
In the APOGEE RC sample, both high- and low-$\alpha$ groups are quite extended in [Fe/H], and the bimodality is seen over
$\sim$0.5 dex of metallicity.  In contrast, the two $\alpha$ groups in the SEGUE studies extend over smaller
ranges in metallicity,
with a quick transition from high- to low-$\alpha$ and no clear bimodality at the metal-poor end.
This difference is likely due to higher uncertainties in the SEGUE data, and their
lack of low-latitude fields (the distribution of \citealt{Cheng12a}, which include some low-latitude SEGUE fields,
looks qualititatively more similar to our RC data).  The APOGEE distribution compares well to those of other
high-resolution studies \citep[e.g.,][]{Fuhrmann11,Bensby14}, most notably the HARPS sample of local stars
\citep[most within $\sim$45 pc;][]{Adibekyan13} as shown by \citet[][see their Figure 9]{Anders14}, although
our sample is $\sim$10 times larger, and extends to significantly larger distances. 

We identify four main qualitative features of our RC [$\alpha$/Fe] vs.\ [Fe/H] diagram, which are labeled
in the schematic (Figure \ref{fig_alphametals_all} right panel):
\begin{enumerate}[1.]
\item $\alpha$ Bimodality:  In the metallicity range $-0.9<$[Fe/H]$<-0.2$ the high and low-$\alpha$
groups are well separated by a less populated chemical region.
\item The two $\alpha$ groups converge at high metallicity, [Fe/H]$\sim+$0.2.  This feature has also been seen
in other recent high-resolution studies \citep[e.g.,][]{Adibekyan13,Ramirez13,Bensby14,Recio-Blanco14,Bergemann14}.
At high metallicities and low [$\alpha$/Fe], where the two groups
merge, it is not completely clear how to associate stars with the two groups and we, therefore, show
this as a hashed red/blue region in the figure.
\item The region between the $\alpha$ groups is not completely empty.  Stars exist there, but at lower
densities than the high/low-$\alpha$ regions.  This statement is still true when examining the selection
function corrected abundances (Figure \ref{fig_selfunc_alphametals}) and the highest $S/N$ stars; which
should have low scatter from abundance uncertainties.
\item The solar-$\alpha$ group extends over $\sim$1 dex of metallicity ($-0.6<$[Fe/H]$<+0.5$) and has a
``banana'' shape.  This shape is seen in previous studies such as the [Ti/Fe] abundances in
\citet[][their Figure 15.]{Bensby14}.
\end{enumerate}

The clear [$\alpha$/Fe] bimodality in the APOGEE data was first shown by \citet{Anders14} using the
first year data, and we confirm this result with our cleaner red clump sample and find that it is not
caused by selection effects (Figure \ref{fig_selfunc_alphametals}).

\subsection{Spatial Variations}

Figure \ref{fig_alphametals_rzbins} shows the [$\alpha$/Fe] vs.\ [Fe/H] diagram for our RC stars
broken into nine regions in R and $|$Z$|$.  The selection effects have a minor effect in
these small spatial regions.  The same general patterns are seen in these panels as were noted
in Figure \ref{fig_alphametals_all} of all stars, except that the relative contributions of
subpopulations vary.  Panels (b), (d) and (e) show the bimodality most prominently.

\begin{figure}[ht!]
\begin{center}
\includegraphics[trim=12mm 8mm 16mm 8mm,clip,angle=0,scale=0.49]{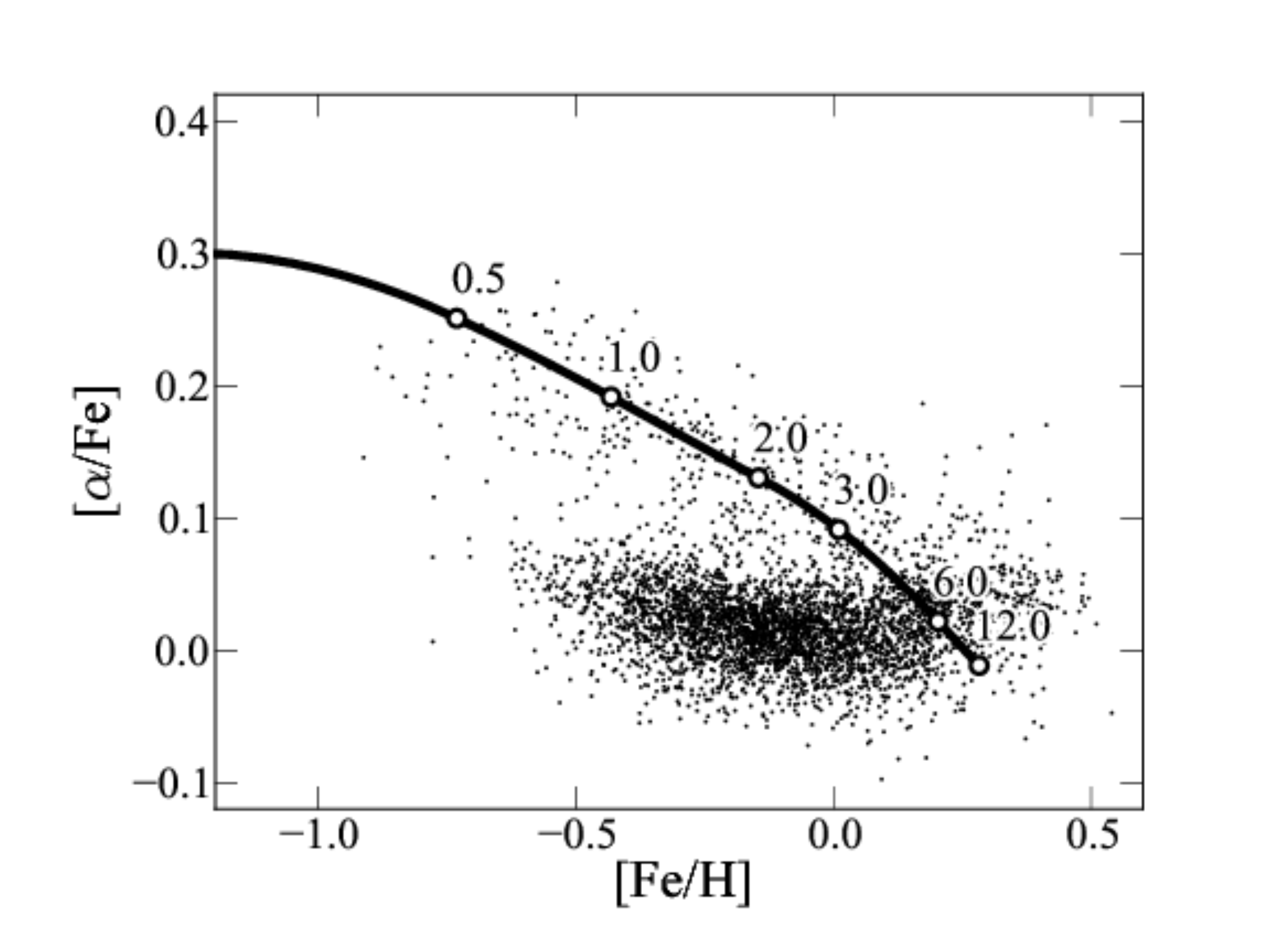}
\end{center}
\caption{The fiducial GCE model for the high-$\alpha$ sequence with SFE=$\sim 4.5 \times 10^{-10}\,{\rm yr}^{-1}$.
The labeled open circles indicate the abundance of each model at the given time in Gyr.}
\label{fig_alphametals_gcemodel_fiducial}
\end{figure}

The high-$\alpha$ stars lie close to the fiducial gray line (in all panels).
Figure \ref{fig_deltaalphahist_rbins} shows $\delta[\alpha$/Fe] (which is [$\alpha$/Fe] minus the fiducial line) for
stars with $S/N$$>$70, $|Z|$$\leq$3 kpc, and $-0.6$$\leq$[Fe/H]$\leq$$-0.2$ in three Galactic radial bins.  A two-Gaussian
fit was performed for each radius, and the means of the Gaussian around $\delta[\alpha$/Fe]=0 are shown as dashed lines.
The means are nearly identical, which suggests that the high-$\alpha$ stars fall along the same trend line
at all radii.  However, when broken out into our nine $R/Z$ zones, some small spatial variations are apparent.
Figure \ref{fig_afeshift_trends} shows the median $\delta[\alpha$/Fe] for high-$\alpha$ stars ([$\alpha$/Fe]$>$$+$0.09)
in the nine $R/Z$ zones. This median $\delta[\alpha$/Fe] displays a negative radial gradient for each zone that increases
in amplitude towards the midplane.  The overall spatial variations are quite small with a RMS of only $\sim$0.015 dex,
or $\sim$10\%.

There are several important qualitative features of Figure \ref{fig_alphametals_rzbins}:

\begin{enumerate}[1.]
\item Constancy of the high-$\alpha$ sequence: The overall shape and position of the high-$\alpha$ sequence does
not vary much with position in the Galaxy, $\sim$10\%.  The small spatial variations show slight negative radial gradients
that increase towards the midplane.

\item The mean metallicity of the low-$\alpha$ sequence changes with $R$:  This reflects the well-known
radial metallicity gradient.  The low-$\alpha$ midplane ($|Z|$$<$0.5) metallicities in the three radial
zones are ($<$[Fe/H]$>$/$\sigma_{\rm [Fe/H]}$):  $+$0.13/0.20, $-$0.01/0.18, $-$0.13/0.19 (inner to outer).
The full RC radial metallicity gradient is presented in \citet{Bovy14}.

\item No high-$\alpha$ sequence for the low-metallicity low-$\alpha$ stars in the outer Galaxy:  There are no
low-metallicity stars with intermediate or high [$\alpha$/Fe] abundances that are linked as a sequence to the
low-metallicity low-$\alpha$ stars at larger radii. This result is in stark contrast to the high-$\alpha$
sequence which connects to the high-metallicity low-$\alpha$ stars (most prominently in the inner Galaxy).

\item The intersection of the high-$\alpha$ and low-$\alpha$ sequences ([Fe/H]$\sim$+0.2) occurs
near the metallicity peak of the low-$\alpha$ sequence in the inner Galaxy,
but well above this mean metallicity in the outer Galaxy (solar annulus
and beyond).  These outer distributions cannot be readily captured by
a single chemical evolution track through [$\alpha$/Fe]-[Fe/H] space.

\end{enumerate}

The implications of these points are discusssed below.

\begin{figure}[ht!]
\begin{center}
\includegraphics[trim=12mm 7mm 16mm 12mm,clip,angle=0,scale=0.49]{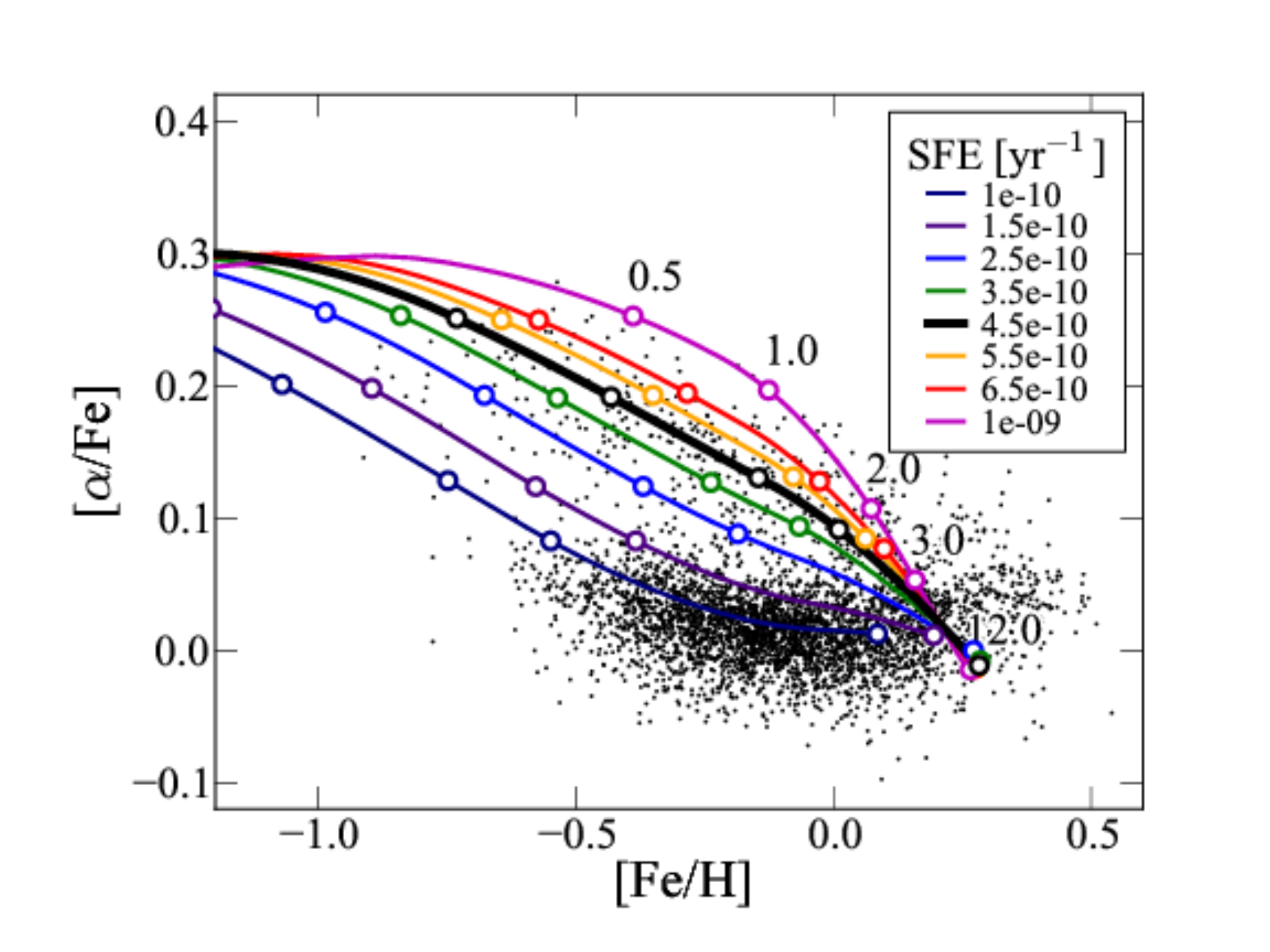} 
\includegraphics[trim=12mm 7mm 16mm 12mm,clip,angle=0,scale=0.49]{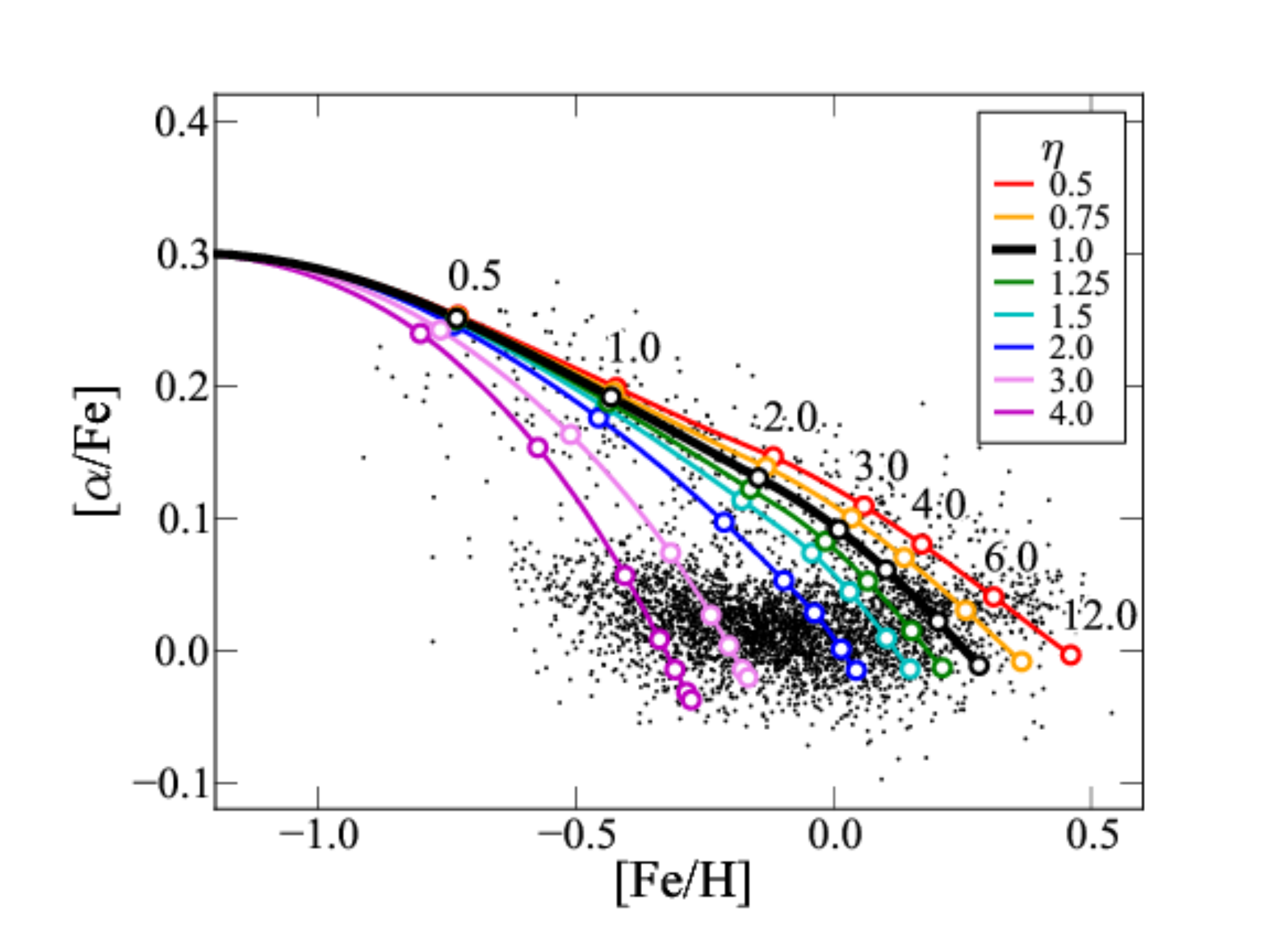}
\includegraphics[trim=12mm 7mm 16mm 12mm,clip,angle=0,scale=0.49]{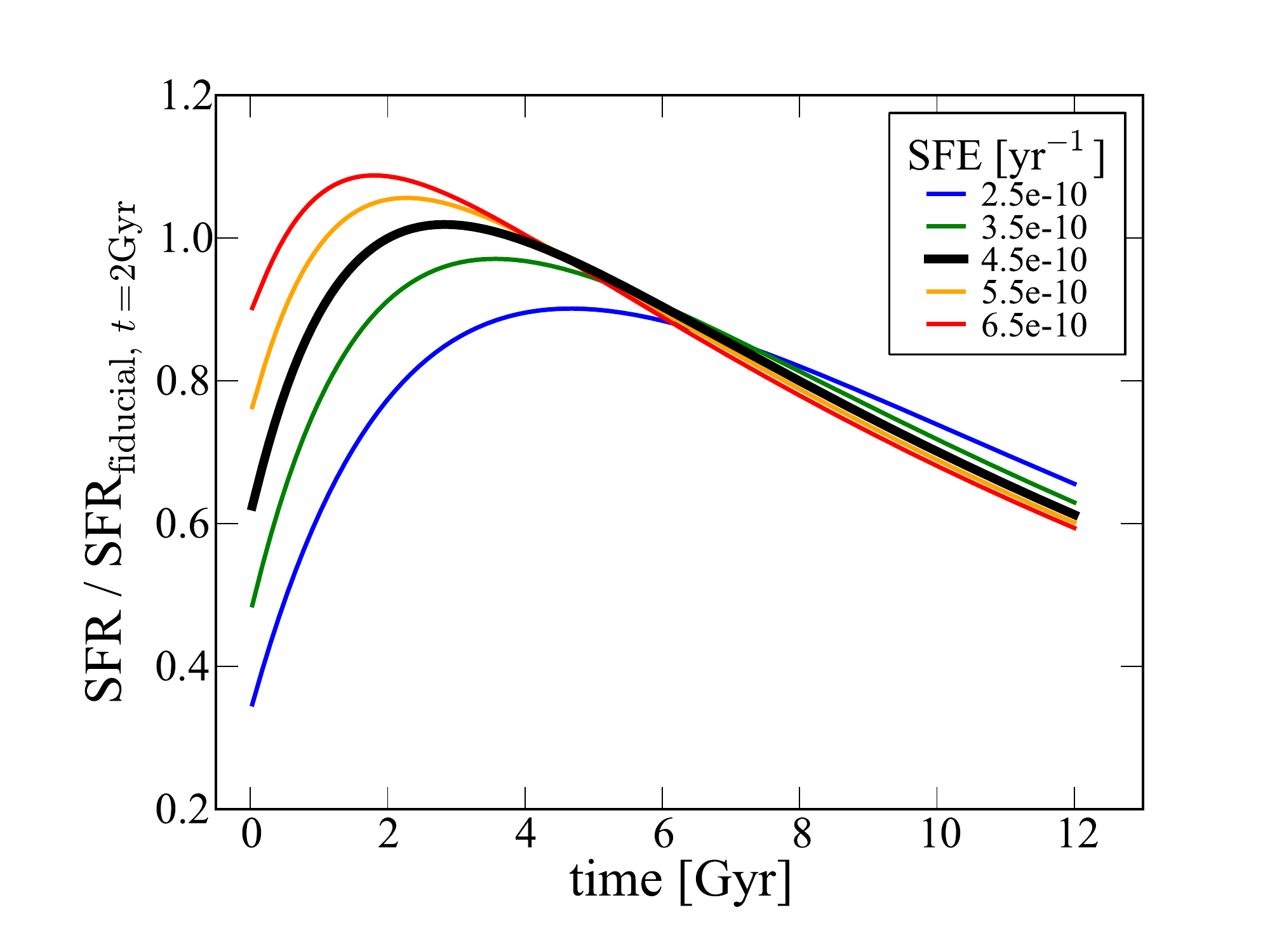}
\end{center}
\caption{Tracks in [$\alpha$/Fe] vs. [Fe/H] for GCE models with
varying star formation efficiency (SFE, top) and outflow rate
($=\eta \times$SFR, middle), as marked in the legend.
The labeled open circles indicate the abundance of each
model at the given time in Gyr.  The APOGEE--RC stars ($S/N$$>$150) are shown as
filled black points.  The heavy black curve indicates the fiducial
model that best fits the observed high-$\alpha$ sequence. (Bottom) The star formation history of the GCE models
normalized by the SFR of the fiducial model at t=2 Gyr.}
\label{fig_alphametals_gcemodel_outflowsfe}
\end{figure}

\section{Discussion}
\label{sec:discussion}

The empirical results presented in \S\ref{sec:results} provide stringent
tests for chemo-dynamical models that can predict the distributions of
iron and $\alpha$-element abundances as a function of Galactocentric radius and
vertical position.  In this section, we examine some of the broad
conclusions that can be drawn from the data, supporting our interpretations
with reference to simple one-zone models of Galactic chemical evolution (GCE).
While these models do not attempt to provide a full account of 
the star formation and enrichment history of the Milky Way, they are
a valuable tool for investigating parameter space to gain a qualitative grasp 
of the effects of the most important processes.

\subsection{Galactic Chemical Evolution Model}
\label{sec:gcemodel}

The GCE model is one-zone with gas accretion and outflow. The inflow rate is
an exponential in time with an $e$-folding time scale of 14 Gyr.  The outflow
rate is parametrized as $\eta \times$SFR, where $\eta$ is the outflow
mass-loading parameter and $\eta = 1.0$ for the fiducial model. We adopt
the yields of \citet{Chieffi04} and \citet{Limongi06} for SNII, the W70 model
of \citet{Iwamoto99} for SNIa, and \citet{Karakas10} for AGB stars.  We assume
a Kroupa (2001) initial mass function (IMF) from 0.1--100 \msune.  The SNIa
delay-time distribution is an exponential in time with an $e$-folding time scale
of 1.5 Gyr and a minimum delay time of 150 Myr.  The star formation rate is
SFE$\times$M$_{\rm gas}$, and we set the star formation efficiency (SFE) to be
4.5$\times$10$^{-10}$ yr$^{-1}$. The GCE model will be described in 
detail by B. Andrews et al. (2014, in preparation), who explore the impact of
varying many different model inputs. For the purposes of this paper, the model
[$\alpha$/Fe] is the average of [O/Fe], [Mg/Fe] and [Si/Fe], which are the
dominant $\alpha$ elements in the APOGEE wavelength range.
This fiducial model is compared to the APOGEE-RC data in Figure \ref{fig_alphametals_gcemodel_fiducial}.

\subsection{The High-$\alpha$ Sequence}

An unexpected feature of the APOGEE--RC $\alpha$-element abundance distributions is
the uniformity of the high-$\alpha$ sequence across
the Galaxy. Figure \ref{fig_alphametals_gcemodel_outflowsfe} shows the
dependence of GCE model tracks in the [$\alpha$/Fe] vs.~[Fe/H] plane on the star
formation efficiency (top) and outflow rate (middle), the two parameters that
have the largest impact on the location of the high-$\alpha$ sequence.  SFE
determines the metallicity of the knee in [$\alpha$/Fe] vs.~[Fe/H] \citep{Pagel97}, and
this in turn sets
the location of the high-$\alpha$ sequence below about solar metallicity.
When the SFE is high, core collapse SNe can enrich the ISM to higher 
[Fe/H] before Type Ia SNe become important and drive [$\alpha$/Fe] toward
solar values.  However, changing the SFE has minimal impact on the 
endpoint of the GCE track, a location in [$\alpha$/Fe] vs. [Fe/H] that
we refer to as the model's ``equilibrium'' abundance, because the evolution
becomes very slow at late times.  This equilibrium abundance is 
sensitive to the outflow rate, since with high $\eta$ the population
cannot evolve to high [Fe/H], because it is losing metals too quickly.
Fitting the observed location of the high-$\alpha$ sequence thus imposes tight
constraints on both SFE and $\eta$, with the former constrained mainly
by the locus below solar [Fe/H], and the latter by the locus above 
solar [Fe/H].

\begin{figure}[ht!]
\begin{center}
\includegraphics[trim=12mm 8mm 16mm 8mm,clip,angle=0,scale=0.49]{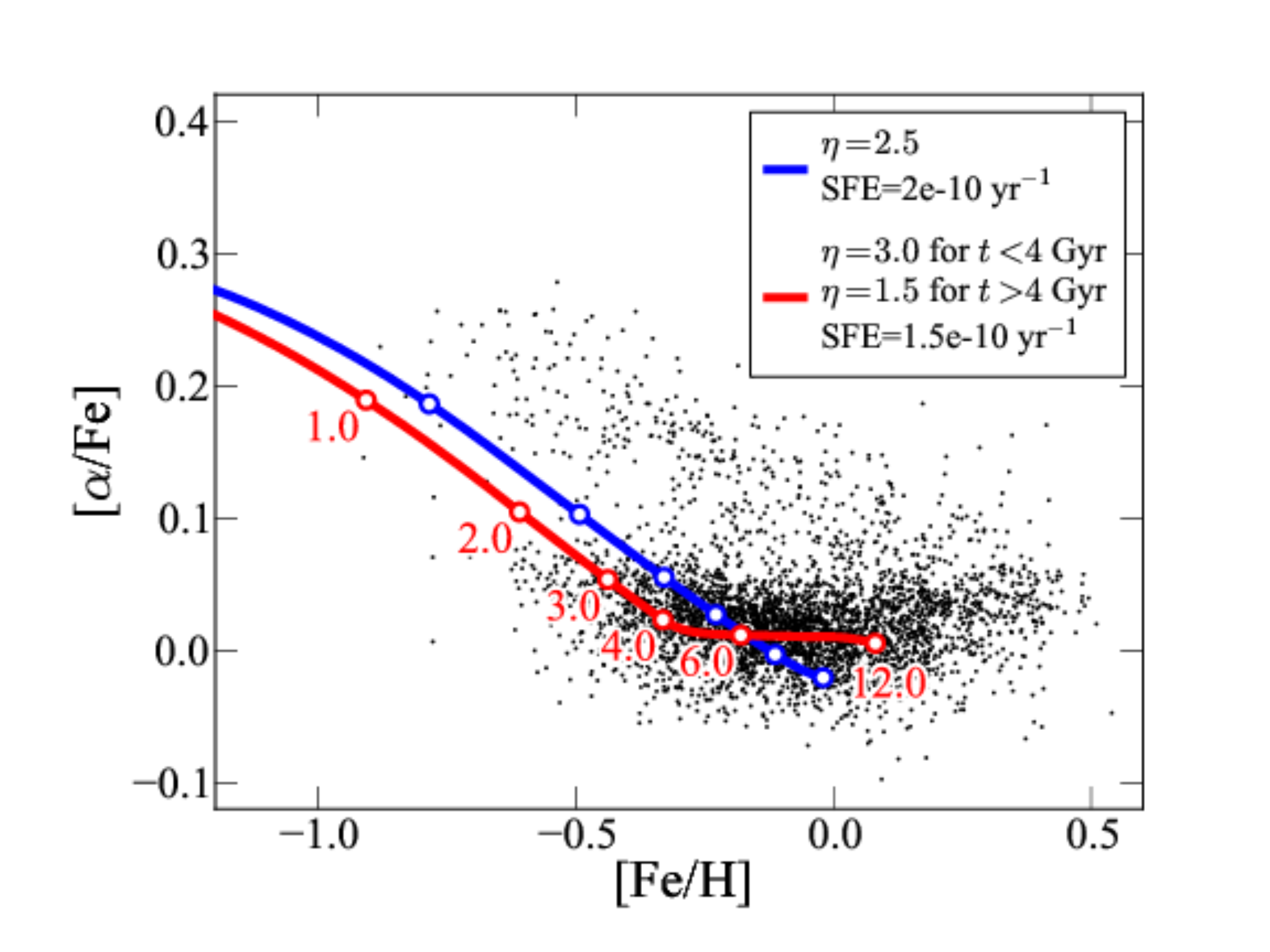}
\end{center}
\caption{Two GCE models intended to reproduce the low-$\alpha$
sequence in the [$\alpha$/Fe] vs.~[Fe/H] plane.  The blue line shows a
model with an outflow rate of 2.5$\times$SFR and SFE=2$\times$10$^{-10}$
yr$^{-1}$, which reaches the density peak of the low-$\alpha$ sequence.  The
red line shows a model with a time-varying outflow rate ($\eta = 3$ for
$t < 4$ Gyr and $\eta = 1.5$ for $t > 4$ Gyr) that runs through the whole
low-$\alpha$ sequence.  The labeled open circles indicate the abundance of
each model at the given time in Gyr.  The APOGEE--RC stars are shown as
filled black points.}
\label{fig_lowalpha_gcemodel}
\end{figure}

The constancy of the high-$\alpha$ sequence in the APOGEE--RC data implies
surprising uniformity in the enrichment history of stars that now occupy a
wide range of $R$ and $|Z|$. It is evident from
Figure~\ref{fig_alphametals_gcemodel_outflowsfe} that plausible model
variations can shift the sequence by amounts much larger than allowed by the
observations. If we treat SFE as the sole adjustable parameter, and fit the
sequence in the nine zones of Figure~\ref{fig_alphametals_rzbins}, we find that the allowed
variations are only $\sim 15\%$. There is some tradeoff between SFE and
$\eta$, and other model inputs to a lesser degree, so more generally this
constancy of location implies a surprising degree of uniformity in some
combination of GCE parameters, with SFE being the most important one.

The implication of a similar SFE for the high-$\alpha$ stars across the Galaxy
appears counter-intuitive because stars in
the inner Galaxy should have formed more rapidly than in the outer Galaxy,
due to higher gas densities.
This in turn would shift the high-$\alpha$
sequence towards higher [Fe/H] at small radii. The SFE for the model that
best reproduces the high-$\alpha$ sequence ($\sim4.5\times10^{-10}\,{\rm
yr}^{-1}$) is remarkably similar to (within the SFE uncertainties) the nearly
constant value ($\sim$5.25$\pm$2.5$\times$10$^{-10}$ yr$^{-1}$; across a diverse
range of local environments) found by \citet{Leroy08} for regions of nearby spiral
galaxies (mostly in the inner parts) dominated by molecular gas (i.e., H$_2$).  This
suggests that the uniformity of the SFE in the early MW could be a product of
an ISM dominated by molecular gas throughout the young, relatively compact disk.
High molecular gas fractions of $\sim$40\% are observed in star forming
galaxies at z$\sim$2 \citep{Tacconi13}, during the era when the MW thick disks are
believed to have formed.
Alternatively, the present distributions could partly be explained if
high-$\alpha$ stars formed predominantly in a narrow, easily-mixed radial annulus in
the inner Galaxy and subsequently migrated to larger radii, creating the more
homogeneous chemical distribution observed today.
It is beyond the scope of this paper to distinguish between these scenarios.
However, observations of disk galaxies at high redshift reveal kinematically hot,
turbulent systems that could contribute to more uniform star formation
conditions across a range of radii (Section \ref{subsec:simulations}).
Therefore, we find it somewhat more likely that the
constancy of the high-$\alpha$ sequence was imprinted at birth, on a stellar
population formed in a well-mixed, turbulent, and molecular-dominated ISM with
a gas consumption timescale (SFE$^{-1}$) of $\sim 2$ Gyr.

\subsection{Low-$\alpha$ Sequence}

Another striking feature of the APOGEE results is the metallicity at which the
high-$\alpha$ sequence merges with the low-$\alpha$ sequence, roughly
[Fe/H]$\approx$+0.2 dex at all locations. In the inner Galaxy this value is
close to the peak of the metallicity distribution function (MDF) of the
low-$\alpha$ population, so it is possible to explain the low-$\alpha$ stars
as the endpoint of the chemical evolution sequence that produced the
high-$\alpha$ population.  However, at the solar radius and in the outer
Galaxy, the merging point of the two $\alpha$ sequences is at a significantly
higher metallicity than the peak of the low-$\alpha$ MDF, by $\sim$0.2--0.3
dex.  This offset makes it difficult (perhaps impossible) to explain the
high-$\alpha$ and low-$\alpha$ stars as the outcome of a single chemical
enrichment history, indicating the presence of at least two distinct
populations.
While this discrepancy has been seen before in the solar neighborhood \citep[e.g.,][]{Bensby11a,Adibekyan13}
the APOGEE RC sample presents the best-populated, most-accurate distributions yet and the first time this
has been plainly seen in the outer Galaxy.

We now describe two different scenarios that could explain the observed chemical
abundances patterns: (1) SFE--transition, and (2) superposition of multiple populations.

\subsubsection{SFE Transition}

Figure~\ref{fig_lowalpha_gcemodel} illustrates two GCE models
with parameters chosen to approximately reproduce the observed locus of the
low-$\alpha$ stars with a single chemical evolutionary sequence. The blue line
shows a model with a low SFE (2$\times$10$^{-10}$ yr$^{-1}$) and high outflow rate (2.5$\times$SFR)
relative to the fiducial model that matches the high-$\alpha$ sequence. The
low SFE shifts the knee of the model sequence to low [Fe/H], and the high
value of $\eta$ shifts the late-time equilibrium abundance to solar [Fe/H].
The model does not produce stars with super-solar [Fe/H]. The red line shows
an alternative model with still lower SFE and an outflow rate that changes
from  $\eta = 3$ for $t < 4$ Gyr to $\eta = 1.5$ for $t > 4$ Gyr. In this
case, the low SFE and high initial $\eta$ drive the model quickly to the low
metallicity end of the low-$\alpha$ locus, but the transition to low $\eta$
allows the model to retain more of the metals that it produces, and thus evolve
to higher [Fe/H] at low [$\alpha$/Fe]. While a decrease in outflow
efficiency at late times is physically plausible, as a result of decreased
star formation rate and a thinner, more settled gas disk, we note that the
parameters of this simulation have been quite finely tuned to match the
location of the observed low-$\alpha$ locus. Small changes in the outflow
mass-loading parameters or the timing of the switch from high to low outflow
rate significantly alter the location of the track in [$\alpha$/Fe]--[Fe/H].

The GCE models can reproduce the observed chemical abundances patterns of the high-$\alpha$
sequence (high SFEH; Figure \ref{fig_alphametals_gcemodel_fiducial}) and the low-$\alpha$ sequence
(low SFE; Figure \ref{fig_lowalpha_gcemodel})\footnote{Note that this is similar to the model suggested by 
\citet{Chiappini09} in which the thin disk (low-$\alpha$ sequence) was formed with a low SFE and
a long timescale infall while the thick disk (high-$\alpha$ sequence) was formed with a high SFE
and a short timescale infall.}.
However, attempts to explain both $\alpha$ groups with a {\it single}
chemical evolution scenario require some special circumstances.  Once the high-$\alpha$ stars reach
low [$\alpha$/Fe] abundance ratios \citep[after $\sim$3--4 Gyr in our GCE model consistent with the
results of][]{Snaith14}, their metallicites are higher than the majority of the younger low-[$\alpha$/Fe]
stars (according to the ages of \citealt{Haywood13}).  Figure 10 of
\citet{Haywood13} demonstrates the difference in metallicity of $\sim$9--10 Gyr old stars between the
high and low-$\alpha$ groups is $\sim$0.5 dex.  To explain the evolution of the low-$\alpha$ stars 
from the gas left over from the formation of the high-$\alpha$ stars (minus the outflow),
the gas would have to be depleted in metals without increasing the overall [$\alpha$/Fe] ratio.
This condition can be accomplished by accretion of large
amounts of pristine gas.
However, if too much gas is accreted too quickly the star formation rate
will rapidly increase and produce many SNII, further increasing the $\alpha$-element abundance ratios to
high values. For example, to decrease the metallicity
from [Fe/H]=$+$0.2 (where the high-$\alpha$ sequence reaches solar-$\alpha$) to [Fe/H]=$-$0.5 (the metal-poor end
of the low-$\alpha$ sequence in the outer Galaxy) requires increasing the gas mass by $\sim$5 times without adding
any metals (i.e., accretion of pristine gas).  From the Kennicutt-Schmidt star formation law 
(\citealt{Kennicutt89}; $\Sigma_{\rm SFR}$ $\propto$ $\Sigma_{\rm gas}^{\rm n}$), this result
implies an increase in the SFR of $\sim$9.5 to 25 with an exponent of n=1.4 and n=2, respectively.  This
enormous rise of the SFR, essentially a ``starburst'', would cause a significant amplification in SNII and
high $\alpha$-element production.
In addition, the large star formation rate would overpredict the observed number of metal-poor stars.
Therefore, the pristine gas has to be accreted on long timescales \citep[e.g.,][]{Chiappini97}.

If the low- and high-$\alpha$ groups are interpreted as two evolutionary sequences (as presented above), then they have
low and high SFE, respectively, and were formed in quite different physical environments.
If the stars that we currently detect in the outer Galaxy all formed
there, the existence of the two $\alpha$-element sequences could represent a dramatic shift in the SFE of the outer
Galaxy $\sim$9 Gyr ago.
In addition to the nearly constant SFE for molecular-dominated ISM, \citet{Leroy08}
found that regions with the ISM dominated by neutral hydrogen gas (\hie) have SFE that decreases with radius.  Since
the outer portions of the MW are now dominated by atomic gas \citep[e.g.,][]{Scoville87,Kulkarni87}, we should expect
the current SFE to decrease with
radius and be low in the outer Galaxy consistent with the observed low SFE low-$\alpha$ sequence.  One possible
scenario is that early on the entire MW gaseous disk was dominated by molecular gas,
producing a nearly uniform SFE and the high-$\alpha$ sequence of stars that we detect.
After $\sim$4 Gyr, the ISM in the outer Galaxy transitioned from molecular-dominated gas (high SFE) to
atomic-dominated gas (lower SFE decreasing with radius), reducing its SFE substantially (by $\sim$1/3), while
the SFE in the inner Galaxy remained high.  This scenario would produce a single high SFE sequence in the inner
Galaxy, but a double SFE sequence in the outer Galaxy.  The transition must have proceeded fairly rapidly to produce the
observed $\alpha$-element bimodality.  The decrease in SFE in the outer Galaxy must have been accompanied by a 
decrease in ISM metallicity to explain the metal-poor low-$\alpha$ stars, thus requiring an infall of pristine gas.
The decrease of SFE combined with a long infall timescale of the pristine gas would help keep the
$\alpha$-abundance ratios low during this active transition period.

\begin{figure}[t]
\begin{center}
\includegraphics[trim=12mm 8mm 16mm 8mm,clip,angle=0,scale=0.49]{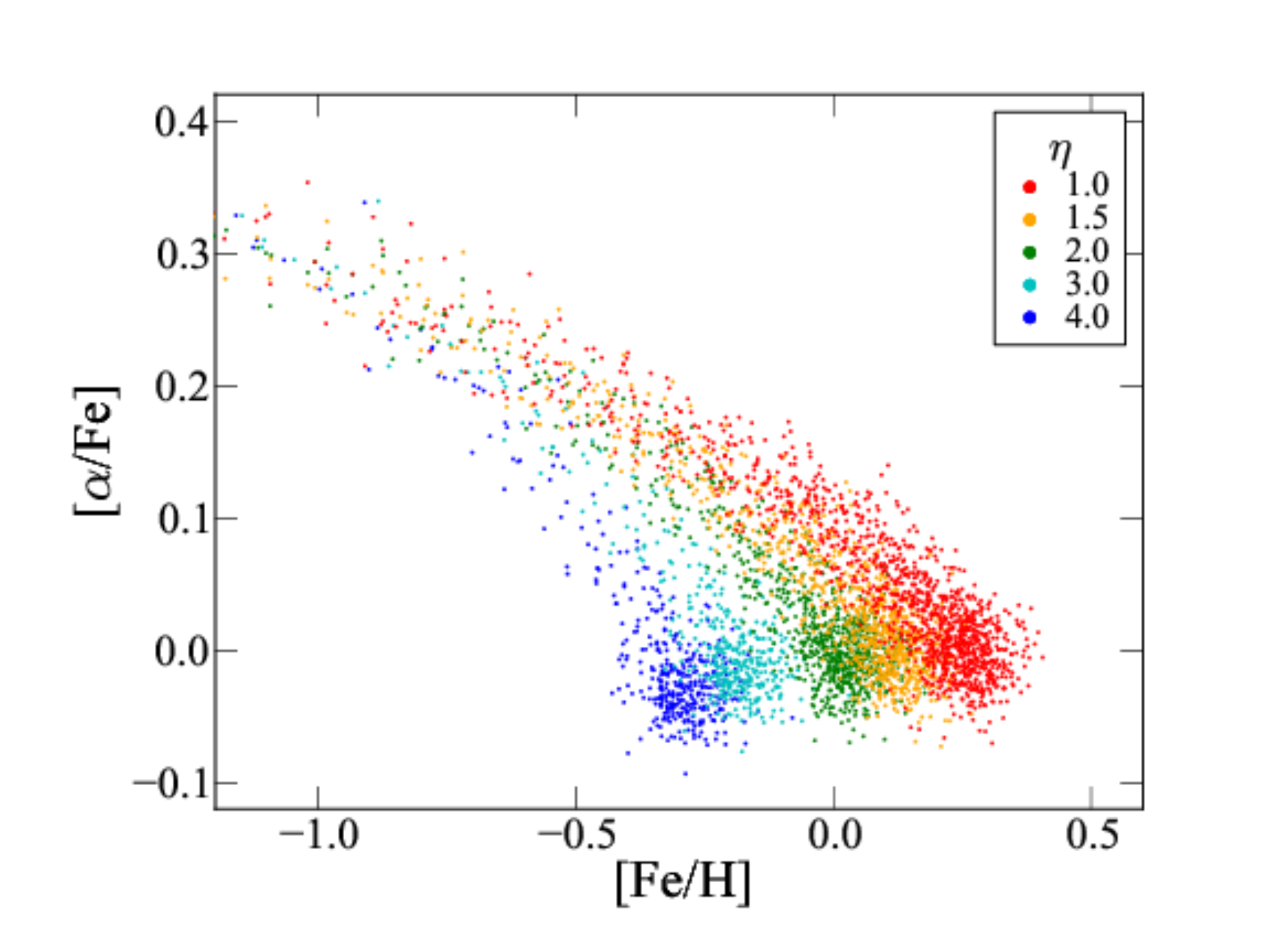}
\end{center}
\caption{A scenario in which the low-$\alpha$ locus
results from the superposition of populations with different enrichment
histories. The different colored points indicate the stellar populations
formed by five different GCE models with outflow rates as marked in the
legend. The models shown by the red, orange, and green points have an
inflow $e$-folding timescale of 14 Gyr, and the blue and cyan models
have a constant SFR.  The red points are from the fiducial high-$\alpha$
model of Figure~\ref{fig_alphametals_gcemodel_outflowsfe}. Gaussian noise
of $\sigma = 0.05$ in [Fe/H] and $\sigma = 0.02$ in [$\alpha$/Fe] was added to
the tracks to show overlapping data points and for ease of comparison with the
data.}
\label{fig_lowalpha_gcemodelsuper}
\end{figure}

While the SFE-transition scenario appears to be a qualitatively
viable interpretation of our results, it also predicts that there should be a fairly rapid transition between
the two SFE values and $\alpha$-element sequences that is not entirely consistent with what is observed.  The \citet{Haywood13}
data (see their Figure 10) indicate a significant overlap in age between the metal-rich end of the high-$\alpha$
sequence and the metal-poor end of the low-$\alpha$ sequence.  While the overlap could partly be explained by
uncertainties in the derived ages, it nevertheless complicates the SFE-transition scenario.

A close inspection of the GCE model for the low-$\alpha$ stars in Figure \ref{fig_lowalpha_gcemodel} indicates
that, while this track fits the
observed low-$\alpha$ sequence fairly well, there is a dearth of observed low-metallicity and intermediate-$\alpha$ stars
in the early part of the track.  These objects are the early ``progenitor'' stars that would have been formed in an ISM
dominated by SNII and high-$\alpha$ abundances and that would eventually produce the SNIa to lower
the $\alpha$-element abundance.  If the low-$\alpha$ sequence of stars formed in ``isolation'', these progenitor stars
should exist, and we should be able to detect them.  These types of stars have recently been found in the Fornax dwarf
spheroidal Galaxy \citep{Hendricks14a}, in an environment with even lower SFR than the outer MW disk. 
Even though our RC sample is biased against metal-poor stars, we should detect stars down to [Fe/H]=$-$0.9, well-below
the observed low-$\alpha$ cutoff of [Fe/H]$\sim$$-$0.5.  Additionally, the lack of progenitor stars for the low-$\alpha$
sequence is evident in independent samples that probe to lower metallicities, such as
\citet[][cutoff at {\rm [Fe/H]}$\ge$$-$0.6]{Fuhrmann11}, \citet[][cutoff at {\rm [Fe/H]}$\ge$$-$0.7]{Adibekyan13},
and \citet[][cutoff at {\rm [Fe/H]}$\ge$$-$0.7]{Bensby14}, which suggests that the paucity of these stars is real.

The lack of progenitor stars of the low-$\alpha$ sequence suggests that the low-$\alpha$ sequence began
its evolution in a low $\alpha$-element abundance ISM.
This scenario could be produced by the stars
in the high-$\alpha$ sequence which formed first polluting the ISM with low-$\alpha$ metals.  This
is a natural consequence in the SFE-transition scenario, since the low-SFE
low-$\alpha$ sequence does not start its evolution with pristine gas, but with the gas chemically enriched
by the high-SFE high-$\alpha$ sequence over $\sim$4 Gyr to low-$\alpha$ abundances.  But, even if the high-$\alpha$
stars formed somewhere else (i.e., in the inner Galaxy), and then moved to their current locations, they could
continuously pollute (pristine accreted) gas in the outer Galaxy at a low level, causing any new
stars formed there to be pre-enriched to a low-$\alpha$ level.  This process essentially causes the low-$\alpha$ chemical
evolutionary sequence to start abruptly at low/intermediate-$\alpha$ and intermediate metallicity ([Fe/H]$\sim$$-$0.7)
which is unexpected from a simple chemical evolutionary analysis.
Additionally, any chemically pristine accreted gas in the outer Galaxy could have been polluted by
low-$\alpha$ outflow from the inner Galaxy.  The \citet{Haywood13} ages indicate that the high-$\alpha$ sequence started
$\sim$13 Gyr ago, while the low-$\alpha$ sequence started $\sim$10 Gyr ago.  By $\sim$10 Gyr ago
the high-$\alpha$ sequence had reached a low enough $\alpha$-element abundance to match those observed for the oldest
low-$\alpha$ stars, albeit at different metallicities (upper panel of Figure 7 of \citealt{Haywood13}).
Not much ($\sim$1/4) intermediate-$\alpha$ metal-rich gas is needed to pollute the pristine gas in the outer
Galaxy to produce the observed metallicity and $\alpha$-element abundance for the oldest stars in the low-$\alpha$ sequence.
This explanation requires continuous injection of enriched gas into the outer Galaxy until the onset of
in situ SNIa's, as well as a fine-tuning of the start of star formation in the outer Galaxy with the chemical
evolution of the inner Galaxy.

\subsubsection{Superposition of Multiple Populations}

The tendency of the models to settle at an equilibrium abundance suggests an
alternative scenario, illustrated in Figure \ref{fig_lowalpha_gcemodelsuper},
in which the low-$\alpha$ locus is not itself an
evolutionary sequence but a superposition of populations that have different
star formation and enrichment histories.   Here we have drawn stars randomly
from the outputs of five models with different outflow rates, one of them (red
points) having the same parameters as the fiducial model used to fit the
high-$\alpha$ sequence, and the other four with higher $\eta$ values that shift
their endpoints to lower [Fe/H]. The three rightmost models (red, orange, and
green points)  have inflow with a 14 Gyr $e$-folding timescale, while the
other two have a constant (instead of declining) SFR, to increase their
equilibrium [$\alpha$/Fe].  All of the other model parameters are the same as
those of the fiducial high-$\alpha$ model.  The final stellar mass of the
low-$\eta$ model is three times that of the highest-$\eta$ models, because of
the differing accretion history and greater retention of ISM gas.  We added
Gaussian noise ($\sigma = 0.05$ dex for [Fe/H] and $\sigma = 0.02$ dex for
[$\alpha$/Fe]) to help visualize the relative numbers of stars with similar
abundances, and compare to the observed locus.  Each model produces the great
majority of its stars close to its endpoint, because early evolution is much
more rapid, and the multiple endpoints merge to form a locus of stars with a
range of [Fe/H] but roughly solar [$\alpha$/Fe]. This picture roughly
resembles the scenario put forward  by \citet{Schoenrich09a}, in which stars
form with different enrichment histories as a function of galactocentric
radius, and the key mechanism for producing a superposition of populations is
radial migration of stars away from their birth radii.

An appealing feature of the superposition scenario is its ability to explain
the shift in the [Fe/H] centroid of the low-$\alpha$ locus with radius.  This
picture requires a low $\eta$ in the inner Galaxy, producing the
high-$\alpha$ sequence and low-$\alpha$, super-solar [Fe/H] stars at its
endpoint, and higher $\eta$ at larger Galactocentric radius to produce tracks
with lower equilibrium [Fe/H]. More efficient outflows could arise in the
outer Galaxy because  of a weaker vertical potential and a lower density of
ISM gas to damp energy injection from supernovae.  Radial gas flows within the
disk can also have a similar effect to radially increasing $\eta$, by
advecting metals produced in the outer disk inward to smaller radii. While the
relative weight of different populations shifts with radius in this scenario,
the APOGEE data show that the high-$\alpha$ sequence is present at all radii,
and is either the relic of an early epoch of star formation or a population formed in
one region that has spread through the Galaxy over time.

An encouraging feature of Figure~\ref{fig_lowalpha_gcemodelsuper} is that it
retains bimodality of the [$\alpha$/Fe] distribution at all metallicities,
with a broader gap at low [Fe/H]. The younger stars of
the high-$\eta$ populations have intermediate values of [$\alpha$/Fe], but the
frequency of these intermediate stars appears at least qualitatively
consistent with the APOGEE data.  The high-$\alpha$ sequence follows the track
of the low-$\eta$ model, in part because it produces the most stars at early
times, but also because the tracks of all five models merge at low [Fe/H],
producing a clear ridge line in the diagram.  There are quantitative
discrepancies between the simulated and observed populations, and in any case
a full model must do more than superpose the results of independent
calculations with tuned parameter choices; it must present a full inflow, star
formation, and outflow history as a function of Galactic position, and specify
whatever mechanisms led to mixing of stellar populations. The comparisons in
Figures~\ref{fig_alphametals_gcemodel_outflowsfe}, \ref{fig_lowalpha_gcemodel},
and~\ref{fig_lowalpha_gcemodelsuper} indicate some of the characteristics that will
be required in a successful model, and they show that the empirical
regularities found in the APOGEE-RC data imply some significant complexities
in the enrichment history of the Milky Way.

\subsection{Comparison to Density Measurements}
This paper has focused on the shape of the high- and low-$\alpha$
sequences and what they imply for the chemical history of the
MW. The relative fraction of stars along each sequence and its spatial
dependence holds important additional clues about how the two
sequences formed and were shaped by evolution. From the radial
dependence of the relative fraction of high- and low-$\alpha$ stars we
conclude that the high-$\alpha$ sequence is primarily associated with
the inner Galaxy, while the low-$\alpha$ stars are most prominent in
the outer Galaxy. This interpretation is in qualitative agreement with the
measurement of the large radial scale length of low-$\alpha$
populations compared to that of high-$\alpha$ populations from SEGUE
\citep{Bovy12b}. Figure \ref{fig_alphametals_rzbins} also suggests that the lowest
$[\mathrm{Fe/H}]$, high-$\alpha$ stars are typically found at larger
$Z$ than the higher $[\mathrm{Fe/H}]$, high-$\alpha$ stars.
This figure shows that this is the case
even at fixed [$\alpha$/Fe].  This result is
also in qualitative agreement with the vertical-scale-height
measurements of \citet{Bovy12b} and
the vertical-metallicity-gradient measurements of \citet{Schlesinger14} and
\citet{Boeche14}.
In future work, we intend to use the
APOGEE data directly to measure the spatial and kinematic
distributions of stars along the high- and low-$\alpha$
sequence. Combined with age dating of the $\alpha$-element sequences, these
measurements will allow us to distinguish between different scenarios
for the origin of the thick disk components in the MW.

\subsection{Simulations and the Extragalactic Context}
\label{subsec:simulations}

The uniform enrichment history of the high-$\alpha$ sequence is predicted by a
simple one-zone model. The physical conditions necessary for one-zone evolution
naturally occur in a thin radial annulus of the young MW disk ($\Delta
r<\sim1$ kpc), where differential rotation and small-scale turbulence can
adequately homogenize the gas-phase metallicity as assumed in many GCE
frameworks \citep[e.g.,][]{Clayton86, Matteucci89}. If born in a small range of
formation radii, the high-$\alpha$ sequence stars must subsequently migrate
throughout the Galaxy to match their currently observed configuration.
Alternatively, large-scale turbulence could mix star-forming gas over longer
distances, widening the formation annulus of the high-$\alpha$ sequence, and
lessening the required degree of radial mixing. While neither scenario can be
ruled out, a well-mixed, globally turbulent young Galaxy is corroborated by
evidence from high-redshift observations. Rotationally-dominated disks observed
at $z\sim2$ exhibit large random motions, and are geometrically thick relative
to local galaxies in both ionized \citep[e.g.,][; and references
therein]{Epinat12,Genzel08} and molecular \citep[e.g.,][]{Swinbank12,
Tacconi13} gas studies. Edge-on UV observations reveal that stellar disks
have similar scale heights to the ionized gas in these systems
\citep[e.g.,][]{Elmegreen06}. If the inferred early dynamical history of the MW
is typical of disk galaxies, our findings offer an important constraint on
galaxy formation models that predict stellar kinematics over a range of
redshift. Numerical simulations in which the early disk forms via a gas-rich
merger \citep{Brook04a}, clumpy star-formation \citep{Bournaud09}, or in situ
star formation from a turbulent star-forming gas reservoir
\citep[e.g.,][]{Bird13}, all qualitatively match the relatively large velocity
dispersions and degree of mixing required by the constant high-$\alpha$
sequence.

The likely one-zone origin of the high-$\alpha$ sequence also suggests that the
MW's radial chemical gradient \emph{was nearly flat} in the past, and
has steepened over time to its current value. There is conflicting
evidence as to the slope of chemical gradients in high redshift
galaxies \citep[e.g.,][]{Yuan13}. Some studies of $z>1$ galaxies report
relatively steep chemical gradients \citep[e.g.,][]{Jones10, Yuan11},
while larger samples suggest that gradients were more shallow than
those found in local spirals \citep{Queyrel12,Swinbank12}. The
temporal evolution of the radial chemical gradient can constrain the
uncertain degree to which energetic feedback mechanisms couple to the
ISM and redistribute metals in simulations of disk galaxy formation
\citep{Gibson13}. Conservative feedback prescriptions predict
initially steep gradients that flatten over time, while models with
'enhanced' feedback physics create flatter gradients at early times
that subsequently steepen \citep{Pilkington12, Gibson13}. Relatively
strong feedback is already required to form disk galaxies with
realistic bulge-to-disk ratios \citep[e.g.,][]{Guedes11} and to
reproduce the stellar mass - halo mass relationship
\citep{Munshi13,Stinson13} found using an abundance-matching approach
\citep[e.g.,][]{Moster13}. Our results are broadly consistent with these
'enhanced' feedback models. The constant high-$\alpha$ sequence suggests
that the chemical radial gradient of the MW has become increasingly
negative with time.

The persistent valley between the low- and high-$\alpha$ sequences is not
readily reproduced by numerical experiments of MW-like galaxies. The
chemical composition distributions of simulated stellar populations
generically show either a single or multiple sequence progression from
low-metallicity, high-$\alpha$ to high-metallicity, low-$\alpha$ regions in the [O/Fe],
[Fe/H] plane \citep[e.g.,][]{Brook12, Minchev13,Stinson13}; others
report relatively flat variation of [O/Fe] with [Fe/H]
\citep[e.g.,][]{Marinacci14}.  While some simulations produce
low-metallicity, low-$\alpha$ stars \citep[e.g.,][]{Roskar13a}, they
fail to reproduce the observed intermediate-$\alpha$ valley.  A variety of
modeling techniques and numerical codes are unable to recreate the
two-dimensional chemical abundance structure observed in the MW.  Therefore,
potential origins of the valley are likely to originate in uncertainties
inherent to all simulations, i.e., the star-formation prescription, feedback
implementations governing mass outflow, and the accretion histories of both
satellite galaxies and gas. Both the spatially-independent high-$\alpha$
sequence and valley are robust structures in the [$\alpha$/Fe], [Fe/H] plane
that offer a promising new avenue of direct comparison with galaxy
formation models.

\section{Conclusions}
\label{sec:conclusions}

We use the APOGEE red clump sample of $\sim$10,000 stars to map the $\alpha$-abundance
patterns across a large volume of the Milky Way disk (5$<$R$<$11 kpc and 0$<$$|$Z$|$$<$2 kpc).
Selection effects for our sample due to the APOGEE targeting strategy and volume probed
are characterized and found not to adversely affect the abundance patterns.  Our main results and
conclusions are as follows:

\begin{enumerate}
\item A bimodality in [$\alpha$/Fe] is detected at low metallicity ($-$0.9$<$[Fe/H]$<$$-$0.2) throughout
the Galaxy. This result is not affected by the
APOGEE targeting and field selection functions.  The low- and high-$\alpha$ abundance sequences merge
at high metallicity ([Fe/H]$\approx$$+$0.2).
\item The shape of the high-$\alpha$ sequence in the [$\alpha$/Fe] vs.\ [Fe/H] diagram is quite constant and
varies little across the Galaxy.  The small spatial variations show a slight negative radial gradient that
increases towards the midplane, but the overall variations are only $\sim$10\%.
The fact that the high-$\alpha$ sequence in the Galactic bulge is similar to the local one \citep{Bensby11b},
in combination with our results implies that the high-$\alpha$ sequence remains nearly constant all the way to the
Galactic center.
\item Using simple galactic chemical evolution models we derive an average star formation efficiency (SFE) in the
high-$\alpha$ sequence of $\sim$4.5$\times$10$^{-10}$ yr$^{-1}$ 
which is quite close to the nearly-constant SFE value for regions of nearby spiral
galaxies dominated by molecular gas.
The homogeneity of the high-$\alpha$ sequence implies that these stars share a similar star formation history and
were formed in a well-mixed, turbulent, and molecular-dominated ISM with a gas consumption timescale (SFE$^{-1}$)
of $\sim 2$ Gyr.
\item The behavior of the high- and low-$\alpha$ sequences as a function of Galactocentric radius shows
that the high-$\alpha$ sequence is more prominent in the inner Galaxy ($R$$\lesssim$$R_0$), while the low-$\alpha$
sequence is more prominent in the outer Galaxy.
\item While the stars in the inner Galaxy can be explained by a single chemical evolutionary track, this cannot
be done in the outer Galaxy, indicating the presence of at least two distinct populations.
\end{enumerate}

A possible expanation for the homogeneity in the shape of the high-$\alpha$ sequence throughout the Galaxy
is that the ISM of the early Milky Way was dominated by molecular gas, which has been shown to have a nearly
constant SFE across diverse physical environments and similar to the value that we measure.  We also
discuss two possible scenarios to explain the low-$\alpha$ sequence (especially in the outer Galaxy):
(1) The gas transitions from high SFE to low SFE coupled with accretion of pristine gas
$\sim$8 Gyr ago.  While this scenario is qualitatively consistent with many of our results, it is not clear
if it agrees in detail with the abundances and ages of \citet{Haywood13} and requires further investigation.
(2) The low-$\alpha$ sequence is composed of a superposition of multiple populations covering a range of outflow
rates and final [Fe/H] on the low-$\alpha$ end.  Mixing of populations formed at different radii could explain
the radial metallicity gradient and the chemical abundance patters in the outer Galaxy.

\acknowledgements

D.L.N. was supported by a McLaughlin Fellowship at the University of Michigan and
thanks Eric Bell, Sarah Loebman, Ian Roederer, Colin Slater, Monica Valluri,
Owain Snaith, Ted Bergin, and Lee Hartmann for useful discussions and suggestions.
J.B. was supported by NASA through Hubble Fellowship grant HST-HF-51285.01 from the
Space Telescope Science Institute, which is operated by the Association of Universities
for Research in Astronomy, Incorporated, under NASA contract NAS5-26555.
J.C.B. acknowledges the support of the Vanderbilt Office of the Provost through the
Vanderbilt Initiative in Data-intensive Astrophysics (VIDA).
B.A., J.J. and D.H.W. acknowledge support from NSF Grant AST-1211853.
M.H., and J.H. acknowledge partial support from NSF Grant AST-1109718,  S.R.M., and A.G.-P.
from AST-1109178, and V.S. from AST-1109888.
TCB acknowledges partial support for this work by grant PHY 08-22648: Physics Frontiers
Center/Joint Institute for Nuclear Astrophysics (JINA), awarded by the U.S. National Science Foundation.
P.M.F. acknowledges support for this research from the National Science Foundation (AST-1311835).
We thank the anonymous referee for useful comments that improved the manuscript.

Funding for SDSS-III has been provided by the Alfred P. Sloan
Foundation, the Participating Institutions, the National Science
Foundation, and the U.S. Department of Energy Office of Science. The
SDSS-III web site is http://www.sdss3.org/.
SDSS-III is managed by the Astrophysical Research Consortium for the
Participating Institutions of the SDSS-III Collaboration including the
University of Arizona, the Brazilian Participation Group, Brookhaven
National Laboratory, Carnegie Mellon University, University of
Florida, the French Participation Group, the German Participation
Group, Harvard University, the Instituto de Astrofisica de Canarias,
the Michigan State/Notre Dame/JINA Participation Group, Johns Hopkins
University, Lawrence Berkeley National Laboratory, Max Planck
Institute for Astrophysics, Max Planck Institute for Extraterrestrial
Physics, New Mexico State University, New York University, Ohio State
University, Pennsylvania State University, University of Portsmouth,
Princeton University, the Spanish Participation Group, University of
Tokyo, University of Utah, Vanderbilt University, University of
Virginia, University of Washington, and Yale University.

This publication makes use of data products from the Wide-field
Infrared Survey Explorer, which is a joint project of the University
of California, Los Angeles, and the Jet Propulsion
Laboratory/California Institute of Technology, funded by the National
Aeronautics and Space Administration.


\end{document}